\documentclass[aps,prd,preprint,groupedaddress,showpacs,showkeys]{revtex4}
\begin{document}
\title{Semiclassical and Quantum Black Holes and their Evaporation, de Sitter and Anti-de Sitter Regimes, Gravitational and String Phase Transitions}
\author{M. Ram\'on Medrano$^{1,2}$ and N. G. S\'anchez$^{1}$}
\email{norma.sanchez@obspm.fr, mrm@fis.ucm.es}
 
\affiliation{
(1) Observatoire de Paris, LERMA, CNRS UMR 8112,  61, Avenue de l'Observatoire, 75014 Paris, France. \\
(2) Departamento de F\'{\i}sica Te\'orica, Facultad de Ciencias F\'{\i}sicas, Universidad Complutense, 
E-28040 Madrid, Spain}

\begin{abstract}

An  effective string theory in physically relevant cosmological and black hole space times is reviewed. Explicit computations of the quantum string entropy, partition function and quantum string emission by  black holes (Schwarzschild, rotating, charged, asymptotically flat, de Sitter $dS$ and $AdS$ space times) in the framework of effective string theory in curved backgrounds provide an amount of new quantum gravity results as: (i) gravitational phase transitions appear with a distinctive {\bf universal} feature: a square root {\bf branch point} singularity in any space time dimensions. This is of the type of the de Vega - S\'anchez transition for the thermal self-gravitating gas of point particles.
(ii) There are no phase transitions in $AdS$ alone. (iii) For $dS$ background, upper bounds of the Hubble constant $H$ are found, dictated by the quantum string phase transition.(iv) The Hawking temperature and the Hagedorn temperature are the same concept but in different (semiclassical and quantum) gravity regimes respectively. (v) The last stage of black hole evaporation is a microscopic string state with a finite string critical temperature which decays as usual quantum strings do in non-thermal pure quantum radiation (no information loss).(vi) New lower string bounds are given for the Kerr-Newman black hole angular momentum and charge, which are entirely different from the upper classical bounds. (vii) Semiclassical gravity states undergo a phase transition into quantum string states of the same system, these states are duals of each other in the precise sense of the usual classical-quantum (wave-particle) duality, which is {\it universal} irrespective of any symmetry or isommetry of the space-time and of the number or the kind of space-time dimensions.

\end{abstract}
\maketitle
{\bf CONTENTS}\bigskip

1. {\bf Introduction and Results}.

\medskip

2. {\bf Semiclassical Backgrounds}:

2.1. Schwarzschild black hole, de Sitter, Anti de Sitter, black hole de Sitter and black hole Anti de Sitter space times.

2. 2. Kerr- Newman (rotating charged) black hole space time.

2. 3. Semiclassical (Q.F.T) Temperature:

2.3.1. Schwarzschild black hole, de Sitter, Anti de Sitter, black hole de Sitter and black hole Anti de Sitter semiclassical temperatures.

2.3.2. Kerr (rotating) black hole and Reissner - Nordstr\"om (charged) black hole semiclassical temperatures. 

2.3.3. Kerr Newman black hole semiclassical temperature.

2.4. (zeroth order) Semiclassical Entropy. 

\medskip

3. {\bf Quantum String Entropy}:

3.1. Entropy for quantum strings in asymptotically flat space times, de Sitter and Anti de Sitter backgrounds. 

3.2. Entropy for quantum strings in Kerr and Reissner - Nordstr\"om black holes space times. 

\medskip

4. {\bf Gravitational String Phase Transitions}:

4.1. Extremal string states and phase transition.

4.2. String phase transition in de Sitter space time.

4.3. Absence of string phase transition in Anti de Sitter space time. 

\medskip

5. {\bf String Partition Function and String Bounds in de Sitter, Anti de Sitter and Black Hole Backgrounds}:

5.1. Partition function and string bounds in de Sitter and Anti de Sitter space times.

5.2. Partition function and string bounds in  black hole backgrounds:

5.2.1. In Schwarzschild black hole space time.

5.2.2. In black hole - de Sitter space time.

5.2.3. In black hole - Anti de Sitter space time.

5.2.4. In Kerr - Newman black hole.

\medskip

6. {\bf Quantum String Emission by a Black Hole and the Last Stage of Black Hole   
    Evaporation} 

6.1. Schwarzschild black hole string emission.

6.2. Black hole - de Sitter string emission.

6.3. Black hole - Anti de Sitter string emission.

6.4. Kerr- Newman black hole string emission.

6.5. Quantum Black Hole Decay.

\medskip

7. {\bf Semiclassical (Q.F.T) and Quantum (String) Regimes}: 

7.1. Semiclassical Entropy for (asymptotically flat) Schwarzschild black hole, de Sitter and Anti de Sitter space times.

7.2. Non extremal Kerr black hole entropy.

7.3. Extremal Kerr black hole and gravitational phase transition.    

7.4. 	De Sitter gravitational phase transition.

7.5. Absence of Anti de Sitter gravitational phase transition.

\medskip

8. {\bf Summary and Conclusions}. \bigskip  

\bigskip

{\bf1. INTRODUCTION AND RESULTS}

\bigskip

The study and solution of the dynamics of strings in curved backgrounds has allowed further understanding of quantum gravity effects \cite{1} - \cite{10}.

In this review paper, classical, semiclassical and quantum regimes are described for cosmological space times in a \emph{conjoint} study. A clear picture for \emph{all} these backgrounds will emerge, going beyond the current picture, both for their semiclassical and quantum regimes. 

The analysis of the semiclassical and quantum regimes of gravity in the contexts of Quantum Field Theory (QFT) and String Theory (ST) respectively is particularly important for several reasons, and it has been the object of recent intensive study, in both regimes, with relevant physical and cosmological examples : Schwarzschild ($BH$)~\cite{11}, Kerr rotating black holes ($KBH$)~\cite{12}, Reissner - Nordstr\"om charged black holes~\cite{12}, Kerr-Newman rotating charged black holes ($KNBH$)~\cite{12}, de Sitter ($dS$)~\cite{13}-~\cite{14} and anti de Sitter ($AdS$) space times~\cite{15}; Schwarzschild -  de Sitter ($bhdS$)~\cite{14} and Schwarzschild - Anti de Sitter ($bhAdS$) backgrounds as well~\cite{15}. 

The physical significance of these space times is well know. The black hole evaporation process has been a challenging problem, related to the so called \emph{loss of information} \cite{16}~\cite{17} and to the issue of the need of a quantum gravity description for its last stages.

The understanding of  semiclassical and quantum gravity $dS$ regimes is particularly important for several reasons.
The flow of consistent cosmological data (cosmic microwave background, large scale structure, and supernovae observations) has placed $dS$, and quasi-$dS$, regimes as a real part of the standard (concordance) cosmological model ~\cite {18}-\cite {24}, for the description of inflation at an early stage of the Universe (semiclassical or QFT regime) and  acceleration at the present time  (classical regime).

Furthermore, $dS$ or quasi $dS$ quantum regimes, besides their conceptual interest, should be relevant in the stage preceeding semiclassical inflation, their asymptotic behaviour should provide consistent initial states for semiclassical inflation, and clarify, for instance, the issue of the dependence of the observable primordial cosmic microwave fluctuations on the initial states of inflation. 

The lack till now of a full conformal invariant description of $dS$ background in string theory \cite{1} should not be considered as an handicap for $dS$ space-time, but as a motivation for going beyond the current scarce physical understanding of string theory. 
Moreover, solving the classical and quantum string dynamics in conformal and non conformal invariant string backgrounds, have shown that the {\bf physics} is the same in the two class of backgrounds: conformal and non conformal. The mathematics is simpler in conformal invariant backgrounds, but the main physics, in particular the string mass spectrum, remains the same,  \cite {1}-\cite{4},\cite{7}.

\medskip

Although there is no relevant cosmological motivation to consider $AdS$ background, the understanding of semiclassical and quantum $AdS$ regimes is relevant as well for several reasons.
$AdS$ regimes are illustrative examples to compare 
and contrast with $dS$ regimes, and they allow to see the effects of a negative cosmological constant.
On the other hand, $AdS$ space time provides asymptotic boundary conditions to Black Hole evaporation, and gives a natural infrared cuttoff to the euclidean path integral formulation for quantum gravity, $AdS$ space time acting as a large space box \cite{25}-\cite{27}.
Also the mathematics is simpler in conformal invariant $AdS$ backgrounds (WZWN models) as compared to the non conformal ones, but the main physics (e.g. the string mass spectrum) continues to be the same~ \cite {1}, ~\cite{3}, ~\cite{5}-~\cite{7}.

\medskip

For strings in  flat space-time, weak string coupling  does not describe any of these backgrounds, but for strings, in any of the full curved backgrounds considered here, these space times are non-perturbative from the beginning. 
We start with the full non perturbative background, so strong coupling effects are present even if no explicit string self interaction is added. Of course, more interactions can be included and explored, but strings in curved backgrounds provide an \emph{effective} framework to deal with strong gravity regimes. 
In the lack of any tractable framework for String Field Theory, (even in the simplest flat space situations), the effective string analogue model or thermodynamical approach implemented in curved backgrounds, provides a suitable framework for this purpose. This approach allows to combine both quantum field theory (QFT) and string theory in curved backgrounds and to go further in the understanding of quantum gravity effects.

In this effective framework, strings are considered as a collection of quantum fields $\Phi_n$ coupled to the curved background, and whose masses $m_n$ are given by the degenerate string mass spectrum in the curved space considered.
Each field $\Phi_n$ appears as many times the string degeneracy of the mass level $\rho_{s} (m)$. Although the fields $\Phi_n$ do not interact among themselves, they do with the background. The mass formula, $m (n)$, and the mass density of states, $\rho_{s} (m)$, are obtained by solving the quantum string (non linear) dynamics in the curved background considered. 

\medskip

In this framework, the semiclassical (Q.F.T) and the Quantum (string) regimes  turn out gravity duals of each other, but in the precise sense of the usual classical - quantum (or wave-particle) duality. This classical - quantum duality does not require the existence of any isometry, in any of the curved backgrounds considered, and neither needs \emph{a priori} any symmetry nor compactified dimensions ~\cite{11}-\cite{15},~\cite{28}-\cite{29}. For instance, the $SL(2,R)$ $WZWN$ string model is a clear and explicit realization of the dual behavior of the quantum string and the semiclassical (Q.F.T) regimes. In this string model, which describes bosonic string theory in a $AdS$ space time in $3$ dimensions, the dual behaviors (semiclassical and quantum string behaviors) are built explicitely in \cite{30}.  

\medskip

A central object in string theory is the microscopic string density of states of mass $m$ in a given background, $\rho_s(m)$; in particular, its high mass behavior that depends on the different curved space-times in which strings propagate (this is the Hagedorn behavior in flat space time~\cite{31}). For a given space-time, this mass behavior grows exponentially; aspects as the number of space-time dimensions, critical dimensions, type of strings, and type of string theory only appear in the two dimensionless numerical coefficients of the exponential growth and its amplitude~\cite{32}.  This is precisely the importance of the density of string states, from which the string intrinsic temperature and entropy emerge, and from which the basic important results of string thermodynamics are derived.
\medskip

The string density of mass levels, $\rho_s (m)$, is derived from the string density of level $n$, $d(n)$, and from the string mass spectrum, $m (n)$, in a given background. 
The string density of levels $d(n)$ is the same in flat and in curved space-time. However,  the mass relation between $m$ and $n$, $m(n)$, and thus the mass density of levels, $\rho_{s} (m)$, depend on the curvature of the background geometry considered,  and they are, in general, different from the flat space-time string mass spectrum and the flat space mass level density.
 The mass formula, $m (n)$, is obtained by solving the quantum string dynamics in \emph{each} space time, for example: $m (n)$ and $\rho_s (m)$ in $dS$ are different from the respective flat space time string mass spectrum and mass level  density~\cite{1},~\cite{5}-\cite{7}.
In general, the formulae $m(n)$ and $\rho_s (m)$ depend on the characteristic lengths in the problem: a classical length $L_{c\ell} $ and a fundamental string length $\ell_s =  \sqrt{\hbar \alpha'/c}$, or equivalently on their respective mass scales: $M_{c\ell} = c^2 L_{c\ell}/G$ (D=4) and $ m_s =\ell_s / \alpha'$. Relevant combinations of them emerge in the mass formula $m(n)$ as the $dS$, and $AdS$, string mass $M_s = L_{c\ell}/ \alpha'$ and string length $L_s = \hbar/ cM_s$. 

\medskip

From the string density of states, $\rho_s (m)$,  the string entropy $S_s(m)$ is obtained for the different space times. An important issue is that phase transitions are found: these can be of the Hagedorn-Carlitz ~\cite{31},~\cite{33} type but with a string critical temperature modified whith respect the flat space Hagedorn temperature, and other hand, other phase transitions are found which are of the type of \emph{gravitational phase transitions} \cite{14}. The latter string transitions show a logarithmic square root branch point type of singularity for the entropy (and a pole singularity for the specific heat); this behavior is similar to the one found for a thermal self-gravitating gas of non relativistic particles (de Vega - S\'anchez transition)~\cite{34},~\cite{35},~\cite{36}, ~\cite{65}. This string behavior is \emph{universal}~\cite{29}, ~\cite{66}i.e. it holds in any number of dimensions, and its origin is gravitational interaction in the presence of temperature (as Jean's instability at finite temperature but with a more complex structure).

\medskip

As a consequence of string phase transitions, string bounds emerge on the relevant semiclassical observables such as the Hawking - Gibbons (semiclassical) temperature~\cite{37},~\cite{38} and the Hubble constant (Sections 3 and 4).

A string temperature $ T_{s}$ appears in $dS$ and $AdS$ backgrounds, that can be higher than the flat space (Hagedorn) string  temperature $t_s $. This happens when, for high masses, the string temperature in a given background is $T_s$, instead of $t_s$. $T_{s}$ turns out to be a critical temperature for $dS$, but not for $AdS$ (~\cite{11} - \cite{15}, ~\cite{28} -\cite{29}); $T_{s}$ is the precise quantum dual of the semiclassical (QFT) temperature scale $T_{sem}=\hbar c /(2\pi k_B L_{c\ell})$~\cite{38}. The two temperatures satisfy: $T_{s} = t_{s}^2~ T_{sem}^{-1}$. 
 
\medskip

In section 2, semiclassical (Gibbons - Hawking) temperature and semiclassical (Bekenstein - Hawking, zeroth order) entropy are introduced, in a systematic way, for $BH$, $KBH$, $KNBH$, $dS$, $AdS$, $bhdS$ and $bhAdS$ space times.

\medskip

In Section 3, the microscopic string density of mass levels $\rho_{s}(m)$ and the full quantum string entropy are given for the different backgrounds. An upper mass bound $M_{s}$ for strings in $dS$ space time appears, but not for $AdS$. The effect of the cosmological constant $\Lambda$ is to reduce the string entropy for a $dS$ background ($\Lambda > 0$) as compared with the one in flat space time; on the contrary, the entropy for string states in an $AdS$ space time ($\Lambda < 0$) will be larger than the string entropy in flat space. The string mode angular momentum $j$ is considered for the string mass density of states $\rho_{s}(m, j)$ and the string entropy $S_{s}(j)$ in a Kerr background. A maximal value appears for $j$ equals to $m^{2}\alpha'c$. The effect of the spin mode $j$ will be to reduce the string entropy as compared when no spin mode has been been taken into account. Similar behavior will have the charge mode $q$ when strings are considered in a Reissner - Nordstr\"om background.

\medskip

As a consequence of string entropy singularities that indicate the appearance of string phase transitions, gravitational like phase transitions are analyzed, in particular, in Section 4. For string states in which $j$ reaches its maximal value, i.e. {\bf extremal string states}, a phase transition takes place at  $T \rightarrow \sqrt{(j/\hbar)}~t_s$ that we call {\bf extremal} transition. A smaller "effective string constant" appears    
 \textit{$\alpha^{'}_j~  \equiv ~ \sqrt{\hbar/j}~ \alpha^{'}$} ~ (and thus a
  {\it higher tension}). In $dS$ space time, the transition occurs at the temperature $T_{s}$ \emph{higher} than the flat space string temperature $t_{s}$, i.e. the Hubble constant $H$ "pushes" the string temperature beyond the flat space time (Hagedorn) value $t_{s}$. $H$ induces a higher "effective string tension" as well. No gravitational string phase transition appears in $AdS$ space time. 

\medskip

Phase transitions show up as well in the thermodynamical behavior of the string partition functions and in the quantum black hole string emissions for different backgrounds.

\medskip

In Section 5, a general study for the string partition functions in the different backgrounds is shown. The partition function, $ln Z$, of a string gas in a curved background with critical temperature  shows a singularity at $T_{sem} \rightarrow t_s$ or $T_s$, for \emph{any dimension} $D$. Namely,
$ln Z$ , for excited and highly excited string gas, shows a single pole temperature singularity (Hagedorn -Carlitz transition) at $t_s$ for (asymptotically flat) Schwarzschild black hole ($BH$), Kerr black hole ($KBH$) and Kerr Newman ($KNBH$) backgrounds. But, for $dS$ and $bhdS$ space times,  $ln Z$ shows a square root branch point singularity at $T_s$ . 

\medskip

In the string phase transition,  $T_{sem} \rightarrow T_{s}$, that takes place for strings in $dS$ background, $H$ reaches a maximum value, $H_{s}$, sustained by the string tension $\alpha'^{-1}$ (and the fundamental constants $\hbar$, $c$).
The partition function has no singular behavior at $T_s$ for massive and highly massive strings in $AdS$ and $bhAdS$ space times, and again no phase transition appears.

\medskip

For black holes with asymptotically flat background, the phase transition occurs at a temperature equals to $t_{s}$. For a black hole with an asymptotically $dS$ background ($bhdS$) phase transition takes place at the temperature $T_{s}$. 
These phase transitions imply, for black hole evaporation process, an upper string bound for the Hawking temperature and lower string bounds for the mass and the horizon radius. Furthermore, for the $bhdS$ case, a relation appears between the horizon radius and the cosmological constant. 

\medskip

For $bhAdS$, $\ln Z$ is mathematically well defined for all temperature. However, we are in the string regime when the semiclassical temperature reaches the string temperature $T_{s}$. Consequences are the existence of a \emph{minimal} $AdS$ classical length or a \emph{maximal} $AdS$ string length. A relation between the horizon radius and the cosmological constant appears as well.

Furthermore, for a $KNBH$, lower string bounds for the black hole angular momentum $J$ and charge $Q$ are found, different from the upper semiclassical bounds.

\medskip

In section 6, the quantum string emission cross section by a black hole is studied. The black hole emission is measured by an observer which is at the asymptotic region, i.e. flat, $dS$ or $AdS$ background.  We also consider the quantum (string) black hole decay and the last stage of the black hole evaporation. 

The quantum string emission cross sections show first the Hawking thermal emission at the semiclassical temperature (semiclassical regime). As evaporation proceeds, the black hole temperature increases and highly massive string states dominate the emission.  The already known phase transitions appear again. For a semiclassical (Hawking - Gibbons) temperature reaching the string temperature ($t_{s}$ or $T_{s}$), the black hole undergoes a string phase transition to a microscopic \emph{string}black hole state, which is a condensed finite energy string state with \emph{finite} string temperature $t_s$ (asymptotically flat) or $T_s$ (asymptotically $dS$), and a size and mass both given in terms of the characteristic string lengths $\ell_{s}$ or $L_s$ (Section 5).
This phase transition, undergone by the emitted strings, represents the non perturbative back reaction effect of the string emission on the black hole. 
We see that the temperature does not become infinite but remains bounded by the string temperature in the asymptotically space time considered. Then, there is not a thermodynamical catastrophe in the last stage of black hole emission as it is the case in the semiclassical black hole approximation (when extrapolated to the last stage).

For a black hole in an asymptotically $AdS$ background, there is no phase transition at 
$T_{sem~bhAdS} = T_{s}$, but we are analogously in the string regime.

Through evaporation and decay, $KNBH$ will loose charge and angular momentum (super radiance like processes) at a higher rate than the loss of mass through thermal radiation.  In general, the last stage of evaporation of a semiclassical $KNBH$ will be a stringy state with (intrinsic) string temperature with zero charge and angular momentum.


At the late stage of black hole evaporation, the black hole decays with a string width $\Gamma_s \sim \alpha' t_s$ or  $\Gamma_s \sim \alpha' T_s$ , into all kind of particles with pure (non mixed) quantum radiation as usual quantum strings do. 

The quantum string black hole state will have a life time $\tau = ( \Gamma_{s})^{-1}$. 
In the effective string framework considered here, there is no loss of information (i.e no paradox at all).

\medskip

In Section 7, by precisely identifying the semiclassical and quantum string mutual dual regimes, new formulae are found for the full semiclassical entropies $S_{sem}$ for $BH$, $KBH$, $dS$ and $AdS$ backgrounds , as functions of the Bekenstein-Hawking entropy $S_{sem}^{(0)}$. 

For a low Hubble constant, $dS$ (and $AdS$) regime is alike to the classical Schwarschild black hole regime , i.e. in this regime the leading term to the entropy is the Bekenstein-Hawking (zeroth order) entropy $S_{sem}^{(0)}(H)$.

But for a large Hubble constant, or high curvature $dS$ regime, $S_{sem}^{(0)}(H)$ is subdominant, and the full entropy $S_{sem}(H)$ is different from the Bekenstein-Hawking entropy . Furthermore, for $H \rightarrow c/\ell_{Pl}$ (being $c/\ell_{Pl}$ the Planck length), a phase transition takes place at $T \rightarrow T_{Pl}$ ( $T_{Pl}$, Planck temperature). This is again a gravitational like transition, i.e a square root branch point singularity at the critical temperature, analogous to the the string gas in $dS$ space time.

For $AdS$, contrary to $dS$, no phase transition occurs at $T \rightarrow t_{Plack}$ ( high curvature, $|\Lambda|^{1/2} \rightarrow c/\ell_{Pl}$, $AdS$ regime).

The semiclassical entropy of a Schwarzschild black hole ($J = 0$) is maximal as compared with the semiclassical entropy for a Kerr black hole ($J \neq 0$). The effect of the angular momentum is to reduce the entropy.

The special case of extremal black holes is clarified. The semiclassical extremal $KNBH$ does not evaporate through Hawking radiation, as the Hawking temperature is zero. 
Also, the string temperature cannot be reached, unless the extremal configuration would be already a stringy state. 
The extremal black hole is, among the black holes states, the most stable configuration, in other words the most classical, or semiclassical, one. Thus if a black hole was not extremal at its origin, it will not be extremal at its end. 
In other words, a $KNBH$ cannot become through quantum decay an extremal black hole, and the extremal black hole cannot be the late state of black hole evaporation. 
In particular, for an extremal $KBH$ ($J = GM^{2}/c$), an {\bf extremal phase transition} occurs  at $T \rightarrow \sqrt{(J/\hbar)}~T_{Pl}$. This entropy singularity is a brach point one, similar to the extremal string transition that was analysed for the extremal string states in Section 4.
 
Section 8 presents the Conclusions. \bigskip


{\bf 2. SEMICLASSICAL BACKGROUNDS}\bigskip


{\bf 2.1. Schwarzschild black hole, de Sitter, Anti de Sitter, black hole de Sitter and black hole Anti de Sitter space times}\bigskip

The D-dimensional space times, for $BH$, $dS$, $AdS$, $bhdS$, and $bhAdS$, are described by the metric (static coordinates)
\begin{equation}
ds^{2}=-a(r) ~c^2~dt^{2} + a^{-1}(r) ~dr^2 + r^2 ~d\Omega_{D-2}^2
 \label{eq:m1}
\end{equation}
where: 
\begin{equation}
a_{BH}(r) =1 - \frac{r_g}{r}~~~,~~~a_{dS}(r) =1- \left(\frac{r}{L_{c\ell}}\right)^2~~~,~~~a_{AdS}(r)=1 + \left(\frac{r}{L_{c\ell}}\right)^2
\label{eq:abSA}
\end{equation}
\begin{equation}
a_{bhdS}(r)=1 - \frac{r_g}{r} - \left(\frac{r}{L_{c\ell}}\right)^2~~~,~~~a_{bhAdS}(r)=1 - \frac{r_g}{r}+ \left(\frac{r}{L_{c\ell}}\right)^2
\label{eq:abhSA}
\end{equation}
and
\begin{equation}
 r_g =  \Bigg(  \frac{16 \pi ~G~ M}{c^2(D-2) ~A_{D-2}}\Bigg)^{\frac{1}{D-3}},~~~~
A_{D-2} =  \frac{2\pi ^{\frac{(D-1)}{2}}}{\Gamma \Big(\frac{(D-1)}{2}\Big)}
\label{eq:BAS}
\end{equation}
$r_g$ is the Schwarzschild gravitational radius, and $L_{c\ell}$ is given by 
\begin{equation}
L_{c\ell} = \sqrt{\frac{(D-2)(D-1)}{2~|\Lambda |}} = \frac{c}{|H|}
\label{eq:Lcl}
\end{equation}
being $H$ the Hubble constant and $\Lambda$ the cosmological constant ($\Lambda > 0$ for $dS$; $\Lambda < 0$ for $AdS$)
\begin{equation}
|\Lambda| = \frac{(D-2)(D-1)}{2}~\frac{H^2}{c^2}
\label{eq:Lan}
\end{equation}
($H^2 > 0 $). The horizons for $BH$ and $dS$ are located at $r_g$ and $L_{c\ell}$ respectively. It is clear that there is no event horizon in $AdS$. Horizons for $bhdS$ and $bhAdS$ will be considered later on. \bigskip

{\bf 2.2. Kerr Newman (rotating charged) black hole space time}\bigskip

A charged rotating black hole of mass $M$, charge $Q$ and angular momentum $J$ is described by the Kerr-Newman geometry ( Boyer-Lindquist coordinates $x^{\mu}=(t, r,\theta,\varphi)$, $D = 4$)
\begin{equation}
ds^{2}=-c^2~(1-\frac{\Pi}{\Sigma})~ dt^{2}- 2~c ~\Pi ~L_{J}\sin^{2}\theta ~dt ~d\varphi +\frac{\Sigma}{\Lambda}~ dr^{2}+\Sigma 
~d\theta^{2}+ B  ~\frac{\sin^{2}\theta}{\Sigma}~ d\varphi^{2}
\label{eq:me}
\end{equation}
where 
\begin{equation}
\Pi= r_{g}~ r - L^{2}_{Q}, ~~~\Sigma=r^{2}+ L_{J}^{2} \cos^{2}\theta, ~~~
~~\Lambda=r^{2} - \Pi + L_{J}^{2},~~~~B= (r^{2} + L_{J}^{2})^{2} - \Lambda  L_{J}^{2} \sin^{2}\theta
\label{eq:pi} 
\end{equation}
\begin{equation}
r_{g}=\frac{2GM}{c^{2}},~~~~L_{J}=\frac{J}{Mc},~~~~L_{Q}=\frac{\sqrt{G}Q}{c^{2}}
\label{eq:L}
\end{equation}
($G$ is the gravitational Newton constant), and $r_{g}$, $L_{J}$, $L_{Q}$ satisfy at the classical level, the inequality :
\begin{equation}
(\frac{r_g}{2})^{2} \geq  L_{Q}^{2} +L_{J}^{2}
\label{eq:bcl}
\end{equation}

This equality holds for the extremal black hole, and Eq.~(\ref{eq:bcl}) shows the classical upper bounds for $J$ and $Q$.  The two horizons are located at $r_{+}$ and $r_{-}$ :
\begin{equation}
r_{\pm}= \frac{r_g}{2} \pm \left((\frac{r_g}{2})^{2} -  L_{Q}^{2} - L_{J}^{2}\right)^{1/2} 
\label{eq:rpm}
\end{equation}\bigskip

{\bf 2.3. Semiclassical (Q.F.T. Hawking) temperature}\bigskip

In the context of Quantum Field Theory (Q.F.T) in curved space time, the semiclassical Gibbons - Hawking temperature (~\cite{37}-~\cite{38}) is given by 
\begin{equation}
T_{sem} = \frac{\hbar c}{2 \pi k_B}~~\frac{1}{\mathcal{L}}
\label{eq:Tsem}
\end{equation}
where $\mathcal{L}$ is the length that measures the \emph{quantum size} in the semiclassical - QFT regime of the studied backgrounds~\cite{11} -\cite{15}. \bigskip

2.3.1. \emph{Schwarzschild black hole, de Sitter, Anti de Sitter, black hole de Sitter and black hole Anti de Sitter semiclassical temperatures}\bigskip

For these backgrounds, the different $\mathcal{L}'s$ are given by
 \begin{equation}
\mathcal{L}_{BH}= 2r_g~~~,~~~\mathcal{L}_{dS}= L_{c\ell}(H)= L_{c\ell}(|\Lambda|)= \mathcal{L}_{AdS}
\label{eq:LBSA}
\end{equation}
\begin{equation}
\mathcal{L}_{bhdS}=2~r_g ~ \Bigg(1 - 2~\Big(\frac{r_g }{L_{c\ell}}\Big)^2~~ \Bigg)^{-1}~~~,~~~\mathcal{L}_{bhAdS}=2~r_g ~ \Bigg(1 + 2~\Big(\frac{r_g }{L_{c\ell}}\Big)^2~~ \Bigg)^{-1}
\label{eq:LbhSA}
\end{equation}
Eq.~(\ref{eq:LBSA}) defines the semiclassical AdS temperature $T_{sem~Ads}$, which is for $AdS$ the analogous of the Hawking temperature for backgrounds with a horizon.  

The black hole surface gravity  $\mathcal{K}$ is defined by
 \begin{equation}
T_{sem} = \frac{\hbar}{2~\pi~k_B~c}~\mathcal{K} 
\label{eq:TK}
\end{equation}
and it can be read from Eqs.~(\ref{eq:LBSA}) and (\ref{eq:LbhSA})  for asymptotically flat, $dS$ and $AdS$ space times.\bigskip

2.3.2. \emph{Kerr (rotating) black hole and Reissner-Nordstr\"om (charged) black hole semiclassical temperatures} \bigskip

According to Eq.~(\ref{eq:Tsem}) we have
\begin{equation}
\mathcal{L}_{KBH}= r_{g} \Delta^{-1} (1 + \Delta)~~~,~~~\mathcal{L}_{RNBH}= \frac{r_g}{2} ~\widehat{\Delta}^{-1} ~(1 + \widehat{\Delta})^2
\label{eq:LBH}
\end{equation}
where
\begin{equation}
\Delta = \sqrt{1- \Bigg(  \frac{2~L_J}{r_g} \Bigg)^2}~~~,~~~ \widehat{\Delta} = \sqrt{1-\Big( \frac{2~L_Q}{r_g}\Big)^2}  
\label{eq:landa}
\end{equation}
($\Delta = 1 = \widehat{\Delta}$ for the Schwarzschild case).\bigskip

2.3.3. \emph{Kerr Newman black hole semiclassical temperature}  \bigskip 

The semiclassical or QFT  black hole temperature (Hawking temperature) is 
\begin{equation}
T_{sem}(J, Q) = \frac{\hbar c}{4 \pi k_{B}}~~ \frac{r_+ - r_-}{r_+^2 + L^2_J} 
\label{eq:TsemJQ}
\end{equation}
which can be rewritten in the general form of Eq.~(\ref{eq:Tsem}), with the length $\mathcal{L}$ now given by
\begin{equation}
\mathcal{L}_{KNBH}\equiv L_{c\ell} (J, Q) = \frac{2~L_{c\ell}}{\delta} \left(1~+~\delta~-~\frac{\nu^{2}}{2}\right)
\label{eq:LJQ}
\end{equation}
where
\begin{equation}
\nu\equiv \frac{2L_{Q}}{r_g}~~~~~~,~~~~~~\mu\equiv\frac{2L_{J}}{r_g}~~~~~~,~~~~~~ \delta\equiv \sqrt{1~-~\nu^{2}~-~\mu^{2}}
\label{eq:numu}
\end{equation}
with $r_g$, $L_{J}$ and  $L_{Q}$ given by Eq.~(\ref{eq:L}).
From Eq.~(\ref{eq:bcl}), we have
\begin{equation}
\nu^{2}~+~\mu^{2}\leq 1
\label{eq:nucu}
\end{equation}
and furthermore
\begin{equation}
\mu^{2}\leq 1~~~~~\text{and}~~~~~\nu^{2}\leq 1
\label{eq:mucu}
\end{equation}
That is, in the semiclassical (Q.F.T) regime, one always has
\begin{equation}
T_{sem}(J, Q) \leq  T_{sem}(J=0=Q)
\label{eq:Tclb}
\end{equation}
which implies
\begin{equation}
L_{c\ell}(J, Q)\geq 2~r_g
\label{eq:Lclb}
\end{equation}
The equality
\begin{equation}
\nu^{2}~+~\mu^{2} = 1
\label{eq:nuex}
\end{equation}
corresponds to the extremal Kerr Newman black hole, $T_{sem}(J, Q)_{extremal} = 0$. \bigskip

 {\bf 2.4. Semiclassical Entropy}\bigskip

The Bekenstein-Hawking (zeroth order) entropy (\cite{39} -\cite{40}) is given by the general expression~\cite{29}

\begin{equation}
S_{sem}^{(0)}= \frac{1}{2}  \frac{\mathcal{M}~c^{2}} {T_{sem}} 
\label{eq:S0sem}
\end{equation}
where $\mathcal{M}$ and $T_{sem}$ depend on the background.

For $BH$, $dS$, $AdS$, $KBH$ and $RNBH$ backgrounds, $\mathcal{M}$ is respectively (\cite{12},~\cite{14},~\cite{15}). 
 \begin{equation}
\mathcal{M}_{BH}= M~~~,~~~\mathcal{M}_{dS}= M_{cl}(H)= M_{cl}(|\Lambda|)= \mathcal{M}_{AdS}
\label{eq:MBSA}
\end{equation}
\begin{equation}
\mathcal{M}_{KBH}\equiv M(J)= M \Delta~~~,~~~\mathcal{M}_{RNBH}= M\widehat{\Delta}
\label{eq:MbhSA}
\end{equation}
where $M$, $J$ and $Q$ are the mass, angular momentum, and charge of the black hole; $\Delta$ and  $\widehat{\Delta}$ are given by
 Eq.~(\ref{eq:landa}). $M_{cl}$ is the (classical) mass scale of $dS$ and $AdS$ space times
\begin{equation}
M_{cl} = \frac{c^2}{G}~L_{c\ell}~~~~~~~~(D=4)
\label{eq:Mcl}
\end{equation}
being $G$ the Newton gravitational constant and $L_{c\ell}$ given by Eq.~(\ref{eq:Lcl}).\bigskip

{\bf 3. QUANTUM STRING ENTROPY}\bigskip

In a given background, the full entropy of quantum strings $S_s$ is related to the microscopic string density $\rho_{s}(m)$, and it is defined by
\begin{equation}
\rho_{s}(m) = e^{\frac{S_{s}(m)}{k_B}}
\label{eq:rhos}
\end{equation}\bigskip

{\bf 3.1. Entropy for quantum strings in asymptotically flat space times, de Sitter and Anti de Sitter backgrounds}\bigskip

In order to derive $\rho_{s}(m)$, we notice that the degeneracy $d_n(n)$ of level $n$ (counting of oscillator states and no spins considered here) is the same in flat and in curved space time. The differences, due to the space-time curvature, will enter through the relation $m=m (n)$ of the mass spectrum. Asymptotically, for high n, the degeneracy $d_n(n)$ behaves universally as
\begin{equation}
d_n(n) = n^{-a'}~ e^{b~ \sqrt n}
\label{eq:d}
\end{equation}
where the constants $a'$ and $b$ depend on the space time dimensions and on the type of the strings. For example, for closed bosonic strings (non compact dimensions)
\begin{equation}
b=2\pi\sqrt{\frac{D-2}{6}}~,~~~~~a'=\frac{D+1}{2}
\label{eq:ba}
\end{equation}

The density $\rho_s(m)$ of mass levels and the level degeneracy $d_n(n)$ satisfy
\begin{equation}
\rho_s(m) ~d \Bigg(\frac{m}{m_s}\Bigg)= d_n(n) ~dn 
\label{eq:rhod}
\end{equation}
where $m_s$ is the fundamental string mass
\begin{equation}
m_s = \sqrt{\frac{\hbar}{\alpha'c}}\equiv~\frac{\ell_s}{\alpha'}
\label{eq:ms}
\end{equation}
being $\alpha'$ the fundamental string constant ($\alpha'^{-1}$ is a mass linear density) and $\ell_{s}$ the fundamental string length.

The mass formula $m(n)$ and the mass density of levels $\rho_s(m)$ are obtained by solving the quantum string dynamics in the curved background considered.
In general, these mass formulae for quantum strings, in any background, can be written as 
\begin{equation}
\big( \frac{m}{m_s}\big)^2 \simeq g(n)
\end{equation}
and  
\begin{equation}
\rho_s(m) \simeq \frac{m}{m_s} \Bigg[ ~\frac{ d_n(n)}{g^{'}(n)} ~\Bigg]_{n=n(m)} 
\label{eq:rod}
\end{equation}
For flat, $dS$ and $AdS$ backgrounds, $g(n)$ can be read from the r.h.s of the following equations (closed bosonic strings)~\cite{1} - \cite{7}
 \begin{eqnarray}
g(n)_{flat}&=& \Bigg( \frac{m}{m_s} \Bigg)^2 \simeq 4 ~n
\label{eq:ga}\\
g(n)_{dS}&=&\Bigg( \frac{m}{m_s} \Bigg)^2 \simeq 4 ~n  \Bigg[1- n \Bigg(\frac{m_s}{M_s} \Bigg)^2 \Bigg] 
\label{eq:gb}\\
g(n)_{AdS}&=&\Bigg( \frac{m}{m_s} \Bigg)^2 \simeq 4 ~n  \Bigg[1+ n \Bigg(\frac{m_s}{M_s} \Bigg)^2 \Bigg] 
\label{eq:gc}
\end{eqnarray}
where $M_s$ is the characteristic string mass in $dS$ ($\Lambda > 0$) and $AdS$ ($\Lambda < 0$) space times
\begin{equation}
M_s = \frac{L_{c\ell}}{\alpha'} = \frac{c}{|H| ~ \alpha'}
\label{eq:Ms}
\end{equation}
Furthermore, $M_s$ defines the quantum string $dS$ ($AdS$) length $L_s$:
\begin{equation}
L_s = \frac{\hbar }{M_s ~c}
\label{eq:Ls}
\end{equation}
and the string $dS$ ($AdS$) temperature $T_s$:
\begin{equation}
T_s = \frac{1}{2\pi k_B} ~ M_s~c^2 = \frac{\hbar c}{2 \pi k_B} ~\frac{1}{L_s}= \frac{1}{2 \pi k_B} ~\frac{c^3}{|H| \alpha'} 
\label{eq:Ts}
\end{equation}
$T_s$ is the critical string temperature only in $dS$ space time, as it is shown in Sec.4, Sec.5 and Sec.6 below.  

From Eqs.(\ref{eq:rod}) and (\ref{eq:ga})-(\ref{eq:gc}), we derive the string mass density of levels in flat, $dS$ and $AdS$ space times~\cite{13} - \cite{15}
 \begin{eqnarray}
\rho(m)_{s, flat}&\equiv& \rho_{s} \Big(\frac{m}{m_s}\Big) \simeq  \Big( \frac{m}{m_s} \Big)^{-a} ~e^{\frac{b}{2} \big( \frac{m}{m_s} \big)}
 \label{eq:roa}\\
\rho(m, H)_{s, dS}&\equiv&\rho_s (m, H)\simeq \Bigg (\frac{m}{\Delta_s M_s}\sqrt{\frac {2}{1 - \Delta_s}}\Bigg)~ 
\Bigg(\frac{M_s}{m_s}\sqrt{\frac{1 - \Delta_s}{2}}\Bigg)^{-a}~ e^{\Big(     \frac{bM_s}{m_s}\sqrt{\frac{1 - \Delta_s}{2}}\Big)} 
\label{eq:rob}\\
\rho(m, |\Lambda|)_{s, AdS}&\equiv&\rho_s (m,|\Lambda| )\simeq \Bigg (\frac{m}{\Delta_s M_s}\sqrt{\frac {2}{\Delta_s - 1 }}\Bigg)~ 
\Bigg(\frac{M_s}{m_s}\sqrt{\frac{\Delta_s - 1 }{2}}\Bigg)^{-a}~ e^{\Big(     \frac{bM_s}{m_s}\sqrt{\frac{\Delta_s - 1 }{2}}\Big)} 
\label{eq:roc}
\end{eqnarray}
where $a~\equiv~2a'~-~1$, and 
\begin{equation}
\Delta_s \equiv\sqrt{1 \mp \Big(\frac{m}{M_s}\Big)^2}
\label{eq:deltaSAS}
\end{equation}
($\mp$ for $dS$ and $AdS$ respectively). We see that $M_s$ is an upper mass bound for the strings in $dS$ background. For strings in flat and $AdS$ space times there is not a mass bound. As the Schwarzschild black space time is asymptotically flat, the asymptotic string mass density of levels will coincide with the one in Minkowski space time~\cite{8},~\cite{9}.

Let us introduce the (zeroth order) string entropy $S^{(0)}_{s}$ in flat space time : 
\begin{equation}
S^{(0)}_{s} =   \frac{1}{2}~b~k_B \Big(\frac{m}{m_s}\Big) =\frac{1}{2}~\frac{m~c^2}{t_s} 
\label{eq:S0c}
\end{equation}
where
\begin{equation}
t_{s} =\frac{1}{b ~k_B}~ m_s~ c^{2} \equiv \frac{\hbar~c}{2\pi~k_B}~\frac{1}{\hat{l_s}}
\label{eq:ts}
\end{equation}
($\hat{l_s}= (b/2\pi)~l_s$) being $t_{s}$ the flat space string temperature.
For flat and (asymptotically flat) $BH$, $dS$ and $AdS$ backgrounds, the density of mass levels can be written in terms of $S^{(0)}_{s}$ as
\begin{eqnarray}
 \rho_s(m)&\simeq&\Big( \frac{S_s^{(0)}}{k_B} \Big)^{-a} ~e^{\big( \frac{S_s^{(0)}}{k_B} \big)}
\label{eq:roaS}\\
\rho_s (m, H)&\simeq& \frac{1}{\Delta_s}\sqrt{\frac{1 + \Delta_s}{2}}~ 
\Bigg(\frac{S_s^{(0)}}{k_B}\sqrt{\frac{2}{1 + \Delta_s}}\Bigg)^{-a}~ e^{\Big(     \frac{S_s^{(0)}}{k_B}\sqrt{\frac{2}{1 + \Delta_s}}\Big)}
\label{eq:robS}\\
\rho_s (m, |\Lambda|)&\simeq& \Bigg (\frac{m}{\Delta_s M_s}\sqrt{\frac {2}{\Delta_s - 1 }}\Bigg)~ 
\Bigg(\frac{M_s}{m_s}\sqrt{\frac{\Delta_s - 1 }{2}}\Bigg)^{-a}~ e^{\Big(     \frac{bM_s}{m_s}\sqrt{\frac{\Delta_s - 1 }{2}}\Big)} 
\label{eq:rocS}
\end{eqnarray}
  
From Eqs.~(\ref{eq:rhos}) and (\ref{eq:roaS}) - (\ref{eq:rocS}), we can read the full string entropy $S_{s}$ in flat and (asymptotically flat) $BH$, $dS$ and $AdS$ backgrounds respectively in terms of $S^{(0)}_{s}$
 \begin{eqnarray}
S_{s, flat}(m)&\equiv&S_s(m) = S_s^{(0)}(m) 
-a~k_B~\ln ~\big(\frac{S_s^{(0)}(m)}{k_B}\big)  
\label{eq:Sa}\\
S_{s, dS}(m, H)&=& \hat {S_s}^{(0)}(m,H) 
-a~k_B~\ln ~\big(\frac{\hat {S_s}^{(0)}(m, H)}{k_B}\big) - k_B ~\ln F(m,H)
\label{eq:Sb}\\
S_{s, AdS}(m,|\Lambda|)&=& \hat {S_s}^{(0)}(m,|\Lambda|) 
-a~k_B~\ln ~\big(\frac{\hat {S_s}^{(0)}(m,|\Lambda|)}{k_B}\big) - k_B ~\ln F(m,|\Lambda|)
\label{eq:Sc}
\end{eqnarray}
where for Eqs.~(\ref{eq:Sb}) and (\ref{eq:Sc}) 
\begin{equation}
\hat {S_s}^{(0)}(m)\equiv S_s^{(0)}\sqrt{f(x)}~~,~~F(m) \equiv \sqrt{(1 \mp 4x^2)f(x)}
\label{eq:SsdsHF}
\end{equation}
($\mp$ for $dS$ and $AdS$ respectively) being $x$ the dimensionless variable  
\begin{equation}
x(m, |H|)\equiv  \frac{1}{2}\Big(\frac{m}{M_s}\Big)= 
\frac{m_s}{b M_s}\frac{S_s^{(0)}}{k_B} 
\label{eq:x}
\end{equation}
and 
\begin{equation}
 f(x)\equiv \frac{2}{1+\Delta_s}
\label{eq:deltasm}
\end{equation}
 where $\Delta_s$ is given by Eq.~(\ref{eq:deltaSAS}) (see Eq.~(\ref{eq:x})).

The entropy $S_{s, dS}(m, H)$ of string states in $dS$ space time is smaller than the string entropy for $H = 0$. The effect of the Hubble constant is to reduce the entropy. On the contrary, the entropy $S_{s, AdS}(m,|\Lambda|)$ of string states in $AdS$ background is larger than the string entropy in flat space. The effect of a negative cosmological constant is to increase the entropy.\bigskip

{\bf 3.2. Entropy for quantum strings in Kerr and Reissner - Nordstr\"om black holes space times}\bigskip

The string entropy $S_s(j)$ in a Kerr background is given, in terms of the string mass density of states $\rho_s(m, j)$, by
\begin{equation}
\rho_s(m,j)= e^{\frac{S_s(m,j)}{k_B}} 
\label{eq:rosj}
\end{equation}
where the string mode angular momentum $j$ is considered. This is a generalization of Eq.~(\ref{eq:rhos}) where no spin was considered. The density of levels $d(n, j)$ of level $n$ and mode $j$ is the same in flat and in curved space-time, and, for large $n$, it is given by ~\cite{42}, \cite{43}
\begin{equation}
d(n, j) \sim n^{-a'} \Delta_s^{-2a'} e^{b \sqrt{n}~  \frac{1 + \Delta_s^2}
{2 \Delta_s }} 
\frac{1}{\cosh^2 \Bigg(  \frac{b}{4} \sqrt{n} ~\frac{(1 - \Delta_s^2)}{\Delta_s} \Bigg)} 
\label{eq:dnj}
\end{equation}
where 
\begin{equation}
\Delta_s(n, j) = \sqrt{1 -  \frac{j}{\hbar  n }}~~,~~~~~~ j\leq \hbar  n
\label{eq:landaj}
\end{equation}

The string mass density of states $\rho_s(m, j)$ and the density of levels $d(n, j)$ are as usual related by
\begin{equation}
\rho_s(\bar{m}, j/\hbar) ~d(\bar{m}) = d(n, j) dn 
\label{eq:rojd} 
\end{equation}
The mass spectrum of strings in asymptotically flat spacetimes, as the black hole background, is the same as the flat spacetime string spectrum (Eq.~(\ref{eq:ga})) ~\cite{11}.
From Eqs.~(\ref{eq:dnj}) and (\ref{eq:rojd}) we have for closed strings
\begin{equation}
\rho_s(\bar{m}, j/\hbar) \sim  \bar{m}^{-a} \Delta_{sj}^{-a-1} e^{ \frac{b}{2} \bar{m} 
\Big (\frac{1 +  \Delta_{sj} ^2}{2 \Delta_{sj}}\Big) }
\frac{1}{\cosh^2 \Bigg(\frac{b}{8} \bar{m} \frac{(1 - \Delta_{sj}^2 )}{\Delta_{sj}} \Bigg)} 
\label{eq:rojc}
\end{equation}
where
\begin{equation}
\Delta_s(m, j) \equiv \Delta_{sj}= \sqrt{1 - \frac{4 j }{m^2 \alpha^{'} c}}~~,~~~4j\leq m^2 \alpha^{'} c 
\label{eq:ba}
\end{equation}

From Eq.~(\ref{eq:rojc}), the asymptotic mass density of states can be written as
\begin{equation}
\rho_s(\bar{m}, j/\hbar) \sim  \rho_s(\bar{m}) \bar{F}(\bar{m}, j/\hbar) 
\label{eq:roF}
\end{equation}
where $\rho_s(\bar{m})$ is the spinless flat mass density of states  (Eq.~(\ref{eq:roa})) 
and 
\begin{equation}
\bar{F}(\bar{m}, j/\hbar) = \Delta_{sj}^{-a-1} e^{ \frac{b}{4} \bar{m} 
\frac{1 +\Delta_{sj}^2}{\Delta_{sj}} }
\frac{1}{\cosh^2 \Bigg(\frac{b}{8} \bar{m} \frac{(1 - \Delta_{sj}^2 )}{\Delta_{sj}} \Bigg)} 
\label{eq:F}
\end{equation}
$\bar{F}(\bar{m}, j/\hbar)$ takes into account the effect of the angular modes $j$, being $ \bar{F}(\bar{m},j=0) = 1$. 

With the help of the zeroth order string entropy $S_s^{(0)}$ (Eq.~(\ref{eq:S0c}) for $j=0$), Eq.~(\ref{eq:rojc}) can be expressed as
\begin{equation}
\rho_s(m, j) \sim \Big(\frac{S_s^{(0)}}{k_B}\Big)^{-a} ~ e^{\Big(\frac{S_s^{(0)}}{k_B}\Big)} ~\bar{F}(S_s^{(0)}, j)
\label{eq:roS0}
\end{equation}
with
\begin{equation}
\bar{F}(S_s^{(0)}, j) = \Delta_{sj}^{-a-1} e^{\frac{S_s^{(0)}}{k_B} \frac{(1 - \Delta_{sj}) ^2}{2 \Delta_{sj}}} 
\frac{1}{\cosh^2 \Bigg(  \frac{S_s^{(0)}}{4 k_B} \frac{(1 - \Delta_{sj}^2)}{\Delta_{sj}} \Bigg)} 
\label{eq:Fcos}
\end{equation}
$ \Delta_{sj}$ now reads
\begin{equation}
\Delta_{sj} = \sqrt{1 - \frac{j}{\hbar} \Big(  \frac{k_B~b}{S_{s}^{(0)}} \Big)^2}
\label{eq:deltaSj}
\end{equation}

Therefore, from Eq.~(\ref{eq:rosj}), the string entropy $S_s(m,j)$ in the Kerr background is given by
\begin{equation}
S_s(m,j) = S_s^{(0)} - a ~k_B~\ln \Big (~\frac{S_s^{(0)}}{k_B}~ \Big)~+~k_B~\ln \bar{F}(S_s^{(0)}, j)      
 \label{eq:SjF}
\end{equation}
That is,
\begin{equation}
S_s(m,j) = \Big(\frac{1 + \Delta_{sj}^2}{2 \Delta_{sj}} \Big) S_s^{(0)} - a~k_B~\ln \Big (~\frac{S_s^{(0)}}{k_B}~ \Big) - 
(a + 1)~k_B~ \ln\Delta_{sj} - 2~k_B~ \ln~\cosh \Big [~  \frac{S_s^{(0)}}{4 k_B}  \frac{(1 - \Delta_{sj}^2)}{\Delta_{sj}}~ \Big] 
\label{eq:SjFcos}
\end{equation}
Notice that the last term $k_B\ln \bar{F}(S_s^{(0)}, j)$ in Eq.~(\ref{eq:SjF}) is enterely due to the angular momentum $j \neq 0$.  
The logarithmic terms have a negative sign. For $j=0$, we recover the flat full entropy expression.
For $\Delta_{sj} \neq 0$, the entropy $S_s(m,j)$ of string states of mass $m$ and mode $j$ is smaller than the string entropy for $j=0$. The effect of the spin is to reduce the entropy. $S_s(m,j)$ is maximal for $j=0$ (ie, for $\Delta_{sj} = 1$).

It is instructive to express $S_s(m,j)$ in terms of the quantity $S_s^{(0)}(m, j)$ for $j\neq 0$ :
\begin{equation}
S_{s}^{(0)} (m,j)  =\frac{1}{2} (1+\Delta_{sj})  S_{s}^{(0)}    
\label{eq:Sosj}
\end{equation}
Then, $S_s(m,j)$ expresses as
\begin{equation}
S_s(m,j) = S_s^{(0)}(m,j) - a ~k_B~\ln \Big (~\frac{S_s^{(0)}(m,j)}{k_B}~ \Big)~+~k_B~\ln F(S_s^{(0)}, j) 
\label{eq:SjFs}
\end{equation}
with
\begin{equation}
F(S_s^{(0)},j)= \Big(\frac{1 + \Delta_{sj}}{2}\Big)^{a}~e^{ \Big(\frac{1 -\Delta_{sj}}{2}\Big) \frac{S_s^{(0)}}{k_B}}  ~~\bar{F}(S_s^{(0)},j) 
\label{eq:Fbar}
\end{equation}
(For $j=0$: $F = \bar{F} = 1$ and $S_s^{(0)}(m,j=0)= S_s^{(0)}$).

Explicitely, in terms of the zero order entropy for $j\neq0$, $S_s^{(0)}(m,j)$:
\begin{equation}
F = \Delta_{sj}^{-1} \Big(\frac{1 + \Delta_{sj}}{2\Delta_{sj}}\Big)^{a}~e^{\Big(\frac{1 -\Delta_{sj}}{1 +\Delta_{sj}}\Big) \frac{S_s^{(0)}(m, j)}{k_B \Delta_{sj}}} 
\frac{1}{\cosh^2 \Bigg(  \frac{S_s^{(0)}(m, j)}{2 k_B} \frac{(1 - \Delta_{sj})}{\Delta_{sj}} \Bigg)} 
\label{eq:Fstring}
\end{equation}
The argument of the last ($\ln\cosh $) term in Eq.~(\ref{eq:SjFcos}) is 
\begin{equation}
x_{j} ~\equiv~ \frac{S_s^{(0)}}{4 k_B}\frac{(1 - \Delta_{sj}^2)}{\Delta_{sj}}~=~\frac{b^2}{4\Delta_{sj}}~
\frac {j}{\hbar}~\frac{k_B}{S_s^{(0)}}~ =~ \frac{b}{4\Delta_{sj}}~ \frac{j}{\hbar}~\frac{m_s}{m}
\label{eq:argx}
\end{equation}

For $j/\hbar < (m/m_s)^2$, and $m \gg m_{s}$, that is for low $j$ and very excited string states, $S_s^{(0)}(m,j)$ is the leading term, but for high $j$, that is $j \rightarrow m^2 \alpha^{'} c$, ie $\Delta_{sj} \rightarrow 0$, the situation is {\it very different} as we will see.

Moreover, Eq.~(\ref{eq:SjFs}) for $S_s(m,j)$ allow us to write in Sec. 7 the whole expression for the semiclassical Kerr black hole entropy $S_{sem}$, as a function of the Bekenstein-Hawking entropy $S_{sem}^{(0)}$.
 
 Similarly, the entropy of strings in a Reissner-Nordstr\"om background is given by
\begin{equation}
\rho_s(m, q)= e^{\frac{S_s(m,q)}{k_B}}  
\label{eq:roqS}
\end{equation}
where $\rho_s(m, q)$ is the string density of states of mass $m$ and charge mode  $q$.

For large $m$, $\rho_s(m, q)$ ~\cite{44} ($D=4$) is given by
\begin{equation}
\rho_s(m, q) \sim  \rho_s(m) ~\exp \Big\{\frac{-q^2}{\hbar c (m/m_s)} \Big\}  
\label{eq:romsq}
\end{equation}

Eqs.~(\ref{eq:roqS}) - (\ref{eq:romsq}) yield: 
\begin{equation}
S_s(m,q)  =  S_s(m) - \frac{k_B}{\hbar c}  \frac{b~q^2}{S_{s}^{(0)}} =  S_s^{(0)} - a ~k_B~\ln S_s^{(0)}
 - \frac{k_B}{\hbar c}  \frac{b~q^2}{S_{s}^{(0)}}
\label{eq:Sqb}
\end{equation}
where $S_s(m)$ is the entropy for $q=0$ Eq.~(\ref{eq:Sa}), and $S_{s}^{(0)}$ its leading term (Eq.~(\ref{eq:S0c})).
  
The string entropy $S_s(m,q)$ of mass $m$ and mode charge $q$ is smaller than the entropy for $q = 0$. As the effect of the spin mode $j$, the effect of the charge $q$ is to reduce the entropy; the $q$-reduction is proportional to $q^{2}$, while the $j$-reduction to the entropy is linear in $j$ plus logarithmic corrections.\bigskip


{\bf 4. GRAVITATIONAL STRING PHASE TRANSITIONS}\bigskip

String phase transitions for black holes and $dS$ space times will be dealt at length in Secs. 5 and 6. But, in this section, we want to stress the appearance of gravitational like phase transitions, similar to the one found for a thermal self-gravitating gas of (non relativistic) particles (de Vega - S\'anchez phase transition)\cite{34}-\cite{36}. \bigskip

{\bf 4.1. Extremal string states and phase transition}\bigskip

If we consider string mode angular momentum $j$ for the asymptotic string mass density of states in flat space time and in a Kerr background as well, we define  {\bf "extremal string states"} the states in which $j$ reaches its {\bf maximal} value, that is $j =  m^2 \alpha^{'} c$.
Then the term $S_s^{(0)}(m,j)$ is minimal:
\begin{equation}
S_{s}^{(0)} (m,j)_{extremal}  = \frac{1}{2} ~S_{s}^{(0)}    
\label{eq:Sosextrem}
\end{equation}

and $S_s(m,j)_{extremal}$ ($(j/\hbar) \rightarrow (m/m_s)^2 $) is dominated by
\begin{equation}
S_s(m,j)_{extremal} =  -(a + 1)~k_B~ \ln~ (~ \sqrt{\frac{2}{T}}\sqrt{T - \Big(\frac{j}{\hbar}\Big)^{1/2} {t_s}}~)~+~O(1)
\label{eq:STextremal}
\end{equation}
($T = m c^2~/~k_B b$). This shows that a {\bf phase transition} takes place at 
$T \rightarrow \sqrt{(j/\hbar)}~t_s$, and we call it {\bf extremal} transition. Notice that this is {\bf not} the usual (Hagedorn/Carlitz) string phase transition occuring for $m \rightarrow \infty $, $T \rightarrow t_s$; although, such transition is also present for $j\neq 0$ since $\rho_{s} (m,j)$ has the same $m \rightarrow \infty $ behavior as $\rho_{s}(m)$.\\
The extremal transition we find here is a {\bf gravitational} like phase transition: the square root {\it branch point} behavior near the transition is analogous to that found in the thermal self-gravitating gas of (non-relativistic) particles (by mean field and Monte Carlo methods). This is also the same behavior found for the microscopic density of states and entropy of strings in de Sitter background (Sec.3). 

A particular new aspect here is that the transition shows up at high angular momentum, (while in the thermal gravitational gas or for strings in $dS$ space time as we shall see, angular momentum is not considered, (although it could be taken into account)). 

Since $j\neq 0$, the extremal transition occurs at a temperature  $ t_{sj}~=~ \sqrt{j/\hbar}~t_s $, {\it higher} than the string temperature $t_{s}$.
That is, angular momentum, which acts in the sense of the string tension, appears in the transition as an "effective string tension" : a smaller 
 \textit{$\alpha^{'}_j~  \equiv ~ \sqrt{\hbar/j}~ \alpha^{'}$} ~ (and thus a
  {\it higher tension}). \bigskip      
 
{\bf 4.2. String phase transition in de Sitter space time}\bigskip

Strings in $dS$ and $AdS$ backgrounds have a very different behavior at string temperature $T_s$ and energy range $M_s$, as we saw from the string entropies in both spaces (Sec.3).  

For $m \sim M_s$, the string entropy in $dS$ (Eq.~(\ref{eq:Sb})) behaves as:
\begin{equation}
S_{s,dS}(m,H)_{m \sim M_s} = k_B \ln \sqrt{\frac{M_s}{~(M_s-m)}}~-k_B\ln~2 ~+~k_B \frac{ b}{\sqrt{2}}~(\frac{M_s}{m_s})~-~ a k_B \ln~(\frac{M_s}{m_s})
\label{eq:SsMsS}
\end{equation}
Or, in terms of temperature : 
\begin{equation}
S_{s,dS}(T,H)_{T \sim T_s} = k_B \ln \sqrt{\frac{T_s}{~(T_s-T)}}~-k_B\ln~2 ~+~k_B \frac{ b}{\sqrt{2}}~(\frac{T_s}{t_s})~-~ a k_B \ln~(\frac{T_s}{t_s})
\label{eq:SsTsS}
\end{equation}
($T = m c^2~/~2 \pi k_B$). We see that a phase transition takes place at $m = M_s$, ie  $T = T_s$. This is again a {\bf gravitational} like phase transition, with a square root {\it branch point} singular behavior near the transition.
This string behavior is {\it universal}~\cite{29}: the logarithmic singularity in the entropy (or pole singularity in the specific heat) holds in any number of dimensions, and its origin is gravitational interaction in the presence of temperature, similar to Jeans's instability at finite temperature but here with a more complex structure.

The transition occurs at the temperature $T_{s}$ (Eq.~(\ref{eq:Ts})) {\it higher} than the (flat space) string temperature $t_{s}$ (Eq.~(\ref{eq:ts})):
\begin{equation}
T_s=  \frac{b}{2\pi}\Bigg(\frac{M_s}{m_s}\Bigg) t_s = \frac{b}{2\pi}\Bigg(\frac{ L_{c\ell}}{\ell_s}\Bigg) t_s 
\label{eq:Tsts}
\end{equation}
 
This is so since in $dS$ background, the flat space string mass $m_s$ (Hagedorn temperature $t_s$) is the scale mass (temperature) in the {\it low} $Hm$ regime. But for high masses, the critical string mass in $dS$ is $M_s$, instead of $m_s$; and the critical string temperature in $dS$ is $T_s$ instead of $t_s$.  In $dS$, $H$ "pushes" the string temperature beyond the flat space (Hagedorn) value $t_s$ .

By analogy with $t_s$, $T_s$ can be expressed as 
\begin{equation}
T_s = \frac{1} {bk_B} \sqrt{\frac{\hbar c^3}{\alpha'_H }}~~~~,~~~~ \alpha^{'}_H = \frac{\hbar}{c}\Big(\frac{2\pi}{b}~\frac{H \alpha'}{c}\Big)^2
\label{eq:TsEff}
\end{equation}
That is, $H$, which acts in the sense of the string tension, induces an "effective string tension" $(\alpha^{'}_H) ^{-1} $ in the transition temperature 
: a smaller  $\alpha^{'}_H$, (and thus a {\it higher tension}). 

The effect of $H$ in the transition is similar to the effect of angular momentum we just shaw. 

 When $m > M_s$, the string does not oscillate (it inflates with the background, and the proper string size is larger than the horizon  \cite{1}, \cite {45}-\cite{47}). The meaning of the string de Sitter phase transition (Eq.(\ref{eq:SsMsS}) or (\ref{eq:SsTsS})) is the following: when the string mass becomes $M_s$, it saturates de Sitter universe, the string size $L_s$ (Compton length for $M_s$) becomes $L_{c\ell}$, and the string becomes   ``{\it classical} '' reflecting the classical properties of the background. Then, $M_s$ is the mass of the background $M_{cl}$ (Eq.~(\ref{eq:Mcl}), with $\alpha '$ instead of $G/c^2$): in other words, for $m \rightarrow M_ s$ the string becomes  the ``{\it background} ''. Conversely, and interestingly enough, string back reaction supports this fact: $M_s$ is the mass of $dS$ background in its string regime, and a de Sitter phase with mass $M_s$ (Eq.~(\ref{eq:Ms})) and temperature $T_s$ (Eq.~(\ref{eq:Ts})) is sustained by strings ~\cite{14}, \cite{29}. ($L_s$, $M_s$, $T_s$) Eqs.~(\ref{eq:Ms})-(\ref{eq:Ts}) are the
{\it intrinsic} size, mass and temperature {\it of} de Sitter background in its string (high H) regime. \bigskip

{\bf 4.3. Absence of string phase transition in Anti de Sitter space time}\bigskip

For strings in $AdS$ background there are not phase transitions at string temperature $T_s$ and energy range $M_s$, but strings will become the background as well. 

From Eq.~(\ref{eq:roc}), we see that for very large $m$ ($m\gg M_{s}$) the leading asymptotic behavior of the $AdS$ string mass density of states
 $\rho_{s,AdS}(m, | \Lambda|)$ grows like $ \sim~ e^{\sqrt{m c^2/|H|\hbar}} $ instead of  $e^{m/m_s}$ as in flat space time (Eq.~(\ref{eq:roa})).
The $AdS$ string mass density of states, $\rho_{s,AdS}(m, | \Lambda|)$, expresses in terms of the typical mass scales in each domain: $m/m_s$ (as in flat space) for low masses, $M_s/m_s =(c/|H|)\sqrt{c/\hbar \alpha'}$ for intermediate masses, $ \sqrt{m M_s}/m_s = c \sqrt{m/\hbar |H|}$ for very high masses, and there is {\bf no} extra singular factor for high masses as it is the case in de Sitter space. Notice that for the very heavy strings, the mass scale turns out to be determined by $\hbar |H|/c^2$ and not by $\sqrt{\hbar/\alpha'c}$. 
The above new features translate into the excited string entropy behavior.

 In fact,  
for intermediate or high masses $m \sim M_s$, and for very high masses  $m\gg M_{s}$ ($M_s$ is no longer a mass bound), the entropy behaves respectively as:
\begin{equation}
S_{s,AdS}(m \sim M_s) = k_B \left(\bar{b}\frac{M_s}{~m_s}\right)~-~ a k_B \ln \left(\frac{M_s}{m_s}\right)
\label{eq:SsMs}
\end{equation}
\begin{equation}
S_{s,AdS}(m\gg M_{s}) = k_B  \left(\frac{b} {m_s} \sqrt{\frac{~m~M_s}{~2}}\right)~-~ \left(\frac{1 + a}{2}\right)~ k_B \ln~ \left(\frac{m}{m_s}\right)~-~ \left(\frac{1 - a}{2}\right)~ k_B \ln\left(\frac{M_s}{m_s}\right)
\label{eq:SsmMs}
\end{equation}
($\bar{b}=b((\sqrt{2}-1)/2   )^{1/2}$). Or, in terms of temperature : 
\begin{equation}
S_{s,AdS}(T \sim T_s) = k_B  \left(\frac{T_s}{t_s}\right)~- ~ a k_B \ln\left(\frac{T_s}{t_s}\right)
\label{eq:SsTs}
\end{equation}
\begin{equation}
S_{s,AdS}(T\gg T_{s}) = k_B \left(\frac{1} {t_s} \sqrt{\frac{~T~T_s}{~2}}\right)~-~ \left(\frac{1 + a}{2}\right)~ k_B \ln~ \left(\frac{T}{t_s}\right)~-~ \left(\frac{1 - a}{2}\right)~ k_B \ln~ \left(\frac{T_s}{t_s}\right)
\label{eq:SstTs}
\end{equation}
($T = m c^2~/~2 \pi k_B$). From Eqs.(\ref {eq:SsMs})-(\ref{eq:SsmMs}) or (\ref{eq:SsTs})-(\ref{eq:SstTs}), we see how the entropy behavior in $(m/m_s)$ - which in $AdS$ background is a low mass or low curvature behavior $(|\Lambda|^{1/2}m\alpha'/c <<1)$ - , does become $M_s/m_s$, and then split into the highly excited entropy behavior $\sqrt{m M_s}/m_s$ in the high mass or high curvature $(|\Lambda|^{1/2}m\alpha'/c >>1)$ regime. Therefore, as $|\Lambda|^{1/2}m$ increases we have the following behavior : $m/m_s ~~ \rightarrow ~~  M_s/m_s ~~ \rightarrow ~~ \sqrt{m M_s}/m_s$. 

Furthermore, there is {\bf no} critical string temperature in $AdS$ background, while we shaw that there exists a critical string temperature in de Sitter background. 
Summing up: The density of states  and entropy do not show any singular behavior at finite mass or temperature in $AdS$ background. 
We will see, in the following sections, that there is a critical temperature in the black hole backgrounds as well. 

However, for $m \sim M_{s}$, the string is as massive as the background, in other words, the string itself becomes the background, or conversely, the background becomes the string. As a consequence, $M_s$ and its corresponding temperature $T_s$ must be truly considered in practice as limiting values for the string mass and string temperature respectively in AdS background.
 
Analogously to the $dS$ case, when the string mass becomes $M_s$,the string size $L_s$ (Compton length for $M_s$) becomes $L_{c\ell}$, the string becomes ``{\it classical} '' reflecting the classical properties of the background, or the background becomes quantum. $M_s$ is the mass of the background $M_{cl}$ (Eq.~(\ref{eq:Mcl})), and for $m \rightarrow M_ s$ the string becomes  the ``{\it background} '' . Conversely, string back reaction supports this fact as well:  $M_s$ is the mass of  AdS background in its string regime, and an Anti de Sitter phase with mass $M_s$ (Eq.~(\ref{eq:Ms})) and temperature $T_s$ (Eq.~(\ref{eq:Ts})) is sustained by strings . ($L_s$, $M_s$, $T_s$) are also the {\it intrinsic} size, mass and temperature of $AdS$ background in its string high $|\Lambda|^{1/2}$ regime~\cite{15}.\bigskip


{\bf 5. STRING PARTITION FUNCTION AND STRING BOUNDS IN DE ~SITTER, ANTI DE SITTER AND BLACK HOLES BACKGROUNDS}\bigskip

The canonical partition function is given by (no string angular momentum $j$ considered here)~\cite{33}
\begin{equation}
\ln Z = \frac{V_{D-1}}{(2\pi)^{D-1}} \int^{\infty}_{m_0} d\Big( \frac{m}{m_s}\Big) \rho_s(m) ~
\int d^{D-1}k ~\ln \Bigg\{  \frac{1 + \exp \Big\{- \beta_{sem} \Big[(m^2 c^4 + \hbar^2 k^2 c^2)^{1/2}\Big] \Big\}}
{1 - \exp \Big\{- \beta_{sem} \Big[(m^2 c^4 + \hbar^2 k^2 c^2)^{1/2}\Big] \Big\}} \Bigg\} 
\label{eq:Zg}
\end{equation}

where supersymmetry has been considered for the sake of generality; $D-1$ is the number of space dimensions; $\rho_s(m)$ is the mass density of states in flat or (asymptotically flat)$BH$ space times, $dS$ and $AdS$ backgrounds ((Eqs.(\ref{eq:roa}) -(\ref{eq:roc})); $\beta_{sem}=(k_B T_{sem})^{-1}$ and $T_{sem}$ is the semiclassical temperature Eq.~(\ref{eq:Tsem}); $m_{0}$ is the lowest mass for which the asymptotic behavior of $\rho_s(m)$ is valid. 

Considering the asymptotic behavior of the Bessel function $K_{\nu}(z)$
\begin{equation}
K_{\nu}(z) \sim \Big( \frac{\pi}{2z}\Big)^2 ~e^{-z}
\label{eq:k}
\end{equation}
and the leading order, $n=1$ (higher excited modes: $\beta_{sem} ~m~c^2 \gg 1$), we have
\begin{equation}
\ln~Z \simeq \frac{2 ~V_{D-1}}{(2\pi )^{\frac{D-1}{2}}}~~ \frac{1}{(\beta_{sem}~~\hbar^2 )^{\frac{D-1}{2}}}~ 
~\int^{\infty}_{m_0} d\Big( \frac{m}{m_s}\Big) ~\rho_s(m) ~
m^{\frac{D-1}{2}}~ e^{-\beta_{sem}mc^2}
\label{eq:Zl} 
\end{equation}
(the factor 2 comes from supersymmetry, as leading contribution is the same for bosonic and fermionic sectors).\bigskip

{\bf 5.1. Partition function and string bounds in de Sitter and Anti de Sitter space times}\bigskip

We know that the mass density of levels $\rho_{s}(m, |\Lambda|)$ has a different behavior from the ones in flat space time, $\rho_{s}(m)$, and in $dS$ background, $\rho_{s}(m, H)$. It is crucial to remark here that $M_s$ (Eq.~(\ref{eq:Ms})) is an upper mass bound for strings only in $dS$ but fixes a mass reference in $AdS$ space time, beyond which $AdS$ string states become highly massive. Let us then analyze $\ln Z$, at significant mass ranges in relation with the string $dS(AdS)$ scale $M_s$. 

For $m \ll M_s$, $\rho_{s}(m, H)$ and $\rho_{s}(m, |\Lambda|)$ have the same leading behavior which is given by the flat  space solution $\rho_{s}(m)$. From Eqs.~(\ref{eq:roa}) and (\ref{eq:Zl}), we have for any D-dimensions
\begin{equation}
(\ln Z)_{m\ll M_s} \sim \frac{2 V_{D-1}}{(2\pi)^{ \frac{D-1}{2}}}~
\frac{(m_s)^{\frac{D-3}{2}}}{(\beta_{sem}~\hbar^2)^{\frac{D-1}{2}}}~~
\frac{1}{(\beta_{sem}-\beta_{s})c^2}~e^{-(\beta_{sem}-\beta_{s})~m_0 c^2}
\label{eq:ZmMs}
\end{equation}
where $\beta_{s}=(k_B~t_s)^{-1}$, $t_s$ is given by Eq.~(\ref{eq:ts}), and $\beta_{sem}=\beta_{semdS}$  or $\beta_{sem}=\beta_{semAdS}$ (Eqs.~(\ref{eq:Tsem}) and (\ref{eq:LBSA}) ).

We see that the canonical partition function, for low $|H| m \ll  c/\alpha'$, shows a pole singularity at $T_{sem}\rightarrow t_{s}$. This transition, near the flat space string temperature $t_s$ and for any space time dimension $D$, is common for the string transition in flat space time (Carlitz/Hagedorn transition~~\cite{31},~\cite{33}) or far from the black hole in the $BH$ space-time, and for the string transition in the low curvature regime of $dS$ and $AdS$ space times. For low temperatures $\beta_{sem}\gg \beta_{s}$, (i.e. $T_{sem}\ll t_{s}$),  we recover the non singular Q.F.T exponential decreasing behavior characterized by $T_{sem}$:
\begin{equation}
\ln Z \simeq V_{D-1} ~\Big(\frac {1}{\hbar^2 \beta_s~ \beta_{sem}} \Big)^{\frac{D-1}{2}}~~ e^{- \beta_{sem}~ m_0 c^2 }
\label{eq:ZTh}
\end{equation}

For $m\sim M_s$, the leading behaviors for the canonical partition function $\ln Z$ for $dS$ and $AdS$ backgrounds are :
 \begin{eqnarray}
(\ln Z)_{m \sim M_s}(dS) &\sim& \frac{V_{D-1}}{\left(\beta_{sem}\hbar c\right)^{D-1}}~~
 \sqrt{\frac{\beta_{sem}-\beta_{sdS}} {\beta_{sem}}}
\label{eq:Za}\\
(\ln Z)_{m\sim M_s}(AdS) &\sim&  \frac {2~V_{D-1}}{(2\pi)^{\frac{D-1}{2}}~ (\hbar c \beta_{sem})^{D-1}}~
 \left(\frac{\beta_s}{\beta_{sAdS}} \right)^{\frac{3-D}{2}}~e^{\frac{1}{\beta_{sAdS}}(\beta_s ~-~
  2\pi \beta_{sem})}
\label{eq:Zb}
\end{eqnarray}
where ~ $ \beta_{sdS} = (k_B T_s)^{-1}= \beta_{sAdS}$, being $T_s$ the string $dS$($AdS$) temperature (Eq.~(\ref{eq:Ts})). 

For strings in $dS$, we rewrite Eq.(\ref{eq:Za}) in terms of the temperature
\begin{equation}
(\ln Z)_{T \sim T_s}(dS)\sim V_{D-1}\left(\frac{k_B T_{sem}}{\hbar c}\right)^{D-1}~~
 \sqrt{1 - \frac{T_{sem}}{T_s}} 
\label{eq:ZhT}
\end{equation}

Eq.(\ref{eq:ZhT}) shows a singular behavior for $T_{sem} \rightarrow T_{s}$ which is general for any space-time dimensions $D$; this is a square root branch point at $T_{sem}= T_s$.  That is, a phase transition takes place for $T_{sem} \rightarrow T_s$, which implies $M_{c\ell}\rightarrow  m_{s}$,  $L_{c\ell}\rightarrow \ell_{s}$ (Eqs. (\ref{eq:ms}) and (\ref{eq:ts})).

Furthermore, we see from Eq.~(\ref{eq:ZhT}) that $T_{sem}$ has to be bounded by $T_s$ $(T_{sem} < T_s)$ . In fact, the low mass spectrum temperature condition $T_{sem}< t_{s}$ (Eq.~(\ref{eq:ZmMs})), and the high mass spectrum condition  $T_{sem}< T_{s}$  (Eq.~(\ref{eq:ZhT})) both imply the following upper bound for the Hubble constant $H$ (Eqs.~(\ref{eq:Lcl}),~(\ref{eq:Tsem}),~(\ref{eq:ms}),~(\ref{eq:Ls}), and~(\ref{eq:Ts})):
\begin{equation}
L_{c\ell} > \ell_s, ~~\text{i.e.}, ~~H < \frac{c}{\ell_s}
\label{eq:Hb}
\end{equation}

In the string phase transition,  $T_{sem} \rightarrow T_{s}$, that takes place for strings in $dS$ background, $H$ reaches a maximum value sustained by the string tension $\alpha'^{-1}$ (and the fundamental constants $\hbar$, $c$ as well)~\cite{14}:
\begin{equation}
H_s = c ~\sqrt{\frac{c}{\alpha'\hbar}}, ~~~~(\text{i.e.},~~\Lambda_s = \frac{1}{2 {\ell_s}^2}(D-1)(D-2))
\label{eq:Hmax}
\end{equation}

The highly excited ($m  \rightarrow M_s$) string gas in $dS$ undergoes a phase transition at high temperature, $T_{sem} \rightarrow T_{s}$, into a condensate stringy state. Eq. (\ref{eq:Hmax}) means that the background itself becames a string state. In Sec. 4, we showed, from the entropy $S_{sdS}$, that precisely at $T = T_s$ $(m = M_s)$ the string of mass $m$ in $dS$ space time undergoes a phase transition at $m = M_s$ and becomes the background itself. 
 
QFT and string back reaction computations support this fact: $dS$ background is an exact solution of the semiclassical Einstein equations with the QFT back reaction of matter fields included~\cite{48} -\cite{58}, as well as a solution of the semiclassical Einstein equations with the string back reaction included \cite{1}: for $T_{sem} \ll T_s$, the curvature $R = R (T_{sem}, T_s)$, yields the QFT semiclassical curvature $R_{sem}$ (low H or semiclassical regime), and for $T_{sem}\rightarrow T_s$ it becomes a string state selfsustained by a string cosmological constant $\Lambda_s$ (Eq.(\ref{eq:Hmax})). The leading term of the de Sitter curvature in the quantum regime is given by $R_s = D \: (D-1)\: c / \ell_s^2$  plus negative corrections in an expansion in powers of ($R_{sem}/R_s$) \cite{13}. The two phases, semiclassical and stringy, are dual of each other in the precise sense of the classical-quantum duality \cite{13}, \cite{28}.

For strings in $AdS$ space time we can consider the $m\gg M_s$ mass range as well,
and the leading behavior for the canonical partition function $\ln Z(AdS)$ is similar to the $m\sim M_s$ behavior:
\begin{equation}
(\ln Z)_{m\gg M_s}(AdS) \sim \frac {2~V_{D-1}}{(2\pi)^{\frac{D-1}{2}} (\hbar \beta_{sem} c)^{D-1}}~
 \left(\frac{\beta_s}{\beta_{sAdS}} \right)^{\frac{3-D}{4}}~ ~~e^{\frac{\beta_s}{\beta_{s AdS}}}~~
e^{- 2\pi\frac{\beta_{sem}}{\beta_{sAdS}}} 
 \label{eq:ZmgMs}
 \end{equation}

We see that the canonical partition function for a gas of strings in a $AdS$ background is defined for {\bf all} temperature.
 The partition function, $\ln Z(AdS)$, for excited and highly excited strings in a $AdS$, does {\bf not} feature any singular behavior at $T_s$ in contrast to the one in flat space time which shows a single pole temperature singularity (Carlitz transition), and to the string partition function in $dS$ space-time which shows a branch point singularity.  There is {\bf no} string phase transition at $T_s$ for massive and highly massive strings in $AdS$ space-time, that is for $|\Lambda|^{1/2}m\alpha'/c \sim 1$, contrary to strings in $dS$ space time. Nevertheless, $M_s$ marks the beginning of the string regime for the $AdS$ case~\cite{15}.  

The results of this subsection will allow us to consider the string regimes of a black hole in $dS$ ($AdS$), or asymptotically $dS$ ($AdS$), backgrounds. This will lead to string bounds for the semiclassical (Hawking-Gibbons) temperature and the black hole radius.\bigskip

{\bf 5.2. Partition function and string bounds in black hole backgrounds}\bigskip 

We consider in this subsection the partition function and bounds for the following cases: The Schwarzschild black hole (asymptotically flat space time), the black hole in a asymptotically $dS$ ($bhdS$) and $AdS$ ($bhAdS$) backgrounds, and the rotating charged black hole ($KNBH$).\bigskip

5.2.1. \emph{In Schwarzschild black hole space time}\bigskip

We have already said that the Schwarzschild black hole space time is asymptotically flat. The partition function is given by  Eq.(\ref{eq:Zg}), with the asymptotic string mass density of levels which coincides with the one in Minkowski space time  (Eq.(\ref{eq:roa})). 

For $T_{sem} \ll t_s$ (Eqs.(\ref{eq:Tsem}), (\ref{eq:LBSA}) and~(\ref{eq:ts})), $lnZ$ is given by Eq.(\ref{eq:ZTh}). For $T_{sem} \rightarrow t_s$, $lnZ$ behavior is reproduced by 
Eq.(\ref{eq:ZmMs}). This singular behavior for $T_{sem} \rightarrow t_s$, and all $D$, is typical of a string system with intrinsic Hagedorn temperature, and indicates a string phase transition at $t_s$ to a condensed finite energy state.

As the definition of the partition function implies the condition $T_{sem} < t_s$ on the semiclassical, or Hawking, temperature, and $T_{sem}$ depends on the black hole mass $M$, or on the horizon $r_g$, this condition provides lower bounds for the mass and the horizon radius (minima mass and radius)~\cite{11}
\begin{equation}
r_{min} = \frac{b}{4\pi}l_s \equiv \frac{\hat{l_s}}{2}
\label{eq:rmin} 
\end{equation}
and
\begin{equation}
M_{min} = \frac{\hat{l_s}}{2l_{Pl}}~m_{Pl} = \frac{b}{8\pi}~\frac{m^2_{Pl}}{m_s}
\label{eq:Mmin}
\end{equation}
where $m_{Pl}$ and $l_{Pl}$ are the Planck mass and length
\begin{equation}
m_{Pl}=\sqrt{\frac{\hbar~c}{G}}~~~~~,~~~~~l_{Pl}=\frac{\hbar}{m_{Pl}~c}
\label{eq:mlP}
\end{equation} 
It is interesting to notice that the effect of quantum string matter is to decrease the black hole mass and horizon, and to increase its temperature: life time becomes shorter~\cite{11}. \bigskip

5.2.2. \emph{In black hole - de Sitter space time}\bigskip 

The black hole-de Sitter ($bhdS$) background tends asymptotically to de Sitter space-time; and, asymptotically i.e far from the black hole,  the string mass density of states is $\rho_s(m,H)$ (Eq.~(\ref{eq:rob})). 
Then, we substitute  $\beta_{sem}$  in  Eq.~(\ref{eq:Zg})  by $\beta_{sem~bhdS}$ ($\beta_{sem~bhdS}= (k_{B}T_{sem~bhdS})^{-1}$). The black hole (Hawking) temperature $T_{sem~bhdS}$ will satisfy: 
$T_{sem~bhdS} <  T_s$, (Eqs.~(\ref{eq:Lcl}),~(\ref{eq:LbhSA}),~(\ref{eq:Ts}),~(\ref{eq:ts}) ), which leads to 
\begin{equation}
\ell_s^2 < L_{c\ell}~L_{bhdS}  
\label{eq:lLbhdS}
\end{equation}
This implies the following condition  
\begin{equation}
H~\Big[1 - 2r_g^2\left(\frac{H}{c}\right)^2\Big] < \frac{2r_{g}c}{\ell_s^2}
\label{eq:HbSS}
\end{equation}

The bound would be saturated for a  gravitational radius satisfying 
\begin{equation}
\left(\frac{r_g}{L_{c\ell}}\right)^{2}~+~r_g \frac{L_{c\ell}}{\ell_s^2}~-~\frac{1}{2} = 0
\label{eq:rgb}
\end{equation}

which yields the physical solution
\begin{equation}
\label{eq:rgs}
r_g = \frac{1}{2}~\frac{L_{c\ell}^3}{\ell_s^2}~\Big[ -  1~+~\sqrt{ 1 + 2\left(\frac{\ell_s}{L_{c\ell}}\right)^4~}~\Big]
\label{eq:requal}
\end{equation}

For $L_{c\ell} \gg\ell_s $ we have
\begin{equation}
r_g \simeq \frac{1}{2}~\frac {\ell_s^2} {L_{c\ell}}~\Big[1 + O(\frac {\ell_s}{L_{c\ell}})^2 ~\Big]~~~, ie.~~~ 2 r_g \simeq \frac {H}{c}\ell_s^2 = L_s
\label{eq:rga}
\end{equation}
and for $L_{c\ell}= \ell_s $ 
\begin{equation}
2r_g = 0.73~\ell_s
\label{eq:rgbx}
\end{equation}

Eq.~(\ref{eq:HbSS}) shows the relation between the  Schwarzschild radius and the cosmological constant (Eq.~(\ref{eq:Lan})) when $T_{sem~bhdS} \rightarrow T_s$ (string regime). We see that a black hole in $dS$ space time allows an intermediate string regime, not present in the Schwarzschild black hole alone ($(H=0)$), since in  the $bhdS$ background there are two characteristic string scales: $ L_s$ and $\ell_s$. We have seen that , in an asymptotically flat space-time, the black hole radius becomes $\ell_s$ in the string regime. In an asymptotically $dS$ space time, when $T_{sem~bhdS}$ reaches $T_s$ , the black hole radius $r_g$ becomes the de Sitter string size $L_s$. If the de Sitter radius $L_{c\ell}$ reaches $L_s$, (which implies $L_{c\ell} = \ell_s$), then $r_g$ becomes determined by the scale $\ell_s$ (Eq.(\ref{eq:ms}))~\cite{14}.\bigskip

5.2.3. \emph{In black hole - Anti de Sitter space time}\bigskip

Analogously to the previous cases,  the black hole-Anti de Sitter ($bhAdS$) background tends asymptotically, far from the black hole, to $AdS$ space-time. 
Asymptotically, far from the black hole, $\rho_s(m,|\Lambda|)$ is the string mass density of states for $bhAdS$ (Eq.~(\ref{eq:roc})). Then, for the partition function of a gas of strings, far from the black hole in the $bhAdS$ space-time, we substitute  $\beta_{sem}$  in  Eq.~(\ref{eq:Zg})  by $\beta_{sem~bhAdS}$,  i.e. by the the black hole temperature in $AdS$ space time $T_{sem~bhAdS}$ (Eq.~(\ref{eq:LbhSA})). 
There is no strict bound $T_{sem~bhAdS} < T_s$ emerging from the string partition function (Eq.~(\ref{eq:Zg})) in the $bhAdS$ case, since the $bhAdS$ string partition function is mathematically well defined for all temperature. However, the regime when the semiclassical temperature $T_{sem~bhAdS}$ reaches the string temperature $T_s$  truly characterizes the string regime of the $bhAdS$ background. 
From Eqs.~(\ref{eq:LbhSA}) and~(\ref{eq:Ts}), the condition  
\begin{equation}
T_{sem~bhAdS} = T_s,  
\label{eq:TsemTs}
\end{equation}
yields :
\begin{equation}
\ell_s^2 = L_{c\ell}~L_{bhAdS} 
\label{eq:lLbhAdS}
\end{equation}
which implies the following equation 
\begin{equation}
\ell_s^2~\Big[1 + 2\left(\frac{r_g}{L_{c\ell}}\right)^2\Big] = 2~r_{g} L_{c\ell}
\label{eq:HbS}
\end{equation}

Thus, the black hole gravitational radius $r_g$ satisfies 
\begin{equation}
\left(\frac{r_g}{L_{c\ell}}\right)^{2}~-~r_g~ \frac{L_{c\ell}}{\ell_s^2}~+~\frac{1}{2} = 0
\label{eq:rgb}
\end{equation}
with the solution:

\begin{equation}
r_{g \pm}~  = ~\frac{1}{2}~\frac{L_{c\ell}^3}{\ell_s^2}~\Big[ ~ 1~ \pm~\sqrt{ 1 - 2\left(\frac{\ell_s}{L_{c\ell}}\right)^4~}~\Big]
\label{eq:rgs}
\end{equation}

Both $r_{g +}$ and $r_{g -}$ are physical roots provided:
\begin{equation}
L_{c\ell}~ \geq 2^{1/4}~\ell_s~\equiv L_{c\ell~min}~=~ 1.189~ \ell_s
\label{eq:rgc}
\end{equation}

That is, there is a {\it minimal} AdS classical length $L_{c\ell~min}$, or a {\it maximal} AdS string length $L_{s~max}$ :  
\begin{equation}
L_s~\leq~ 2^{-1/4} \ell_s~\equiv ~L_{s~max} ~=~ 0.841~\ell_s.
\label{eq:rgams}
\end{equation}

For $L_{c\ell} \gg L_{c\ell~min}$, $r_{g +}$ and $r_{g -}$ are : 
\begin{equation}
r_{g +} ~ =~ \frac {L_{c\ell}^3} {\ell_s^2}~\Big[1 - \frac{1}{2}\left(\frac {\ell_s}{L_{c\ell}}\right)^4 + O\left(\frac {\ell_s}{L_{c\ell}}\right)^8 ~\Big] 
\label{eq:rga}
\end{equation}
\begin{equation}
r_{g -} ~=~ \frac {1}{2}~\frac{\ell_s^2} {L_{c\ell}}~\Big[~1 - O\left(\frac {\ell_s}{L_{c\ell}}\right)^2 ~\Big] 
\label{eq:rgg}
\end{equation}
which in terms of $L_s$  read: 
\begin{equation}
r_{g +} ~ \simeq \frac {L_{c\ell}^2}{L_s}=\frac {c^4}{\hbar\alpha'H^3}~~~~,~~~~r_{g -} \simeq \frac {1}{2}~ L_s= \frac {\hbar\alpha'H}{2c^2}~~~~,~~~{\text ie}~~~  ~r_{g +}r_{g-}= \frac{L_{c\ell}^2}{2} 
\label{eq:rgam}
\end{equation}
 
On the other hand, for  $L_{c\ell}~ =~ L_{c\ell~min}~=~ 2^{1/4}~~\ell_s$ (or $L_s~ =~ L_{s~max}~=~ 2^{-1/4}~~\ell_s$), and $r_{g~ \pm}$ are:
\begin{equation}
r_{g +}~ = ~ r_{g -}~ = ~ \frac {L_{c\ell~min}}{\sqrt{2}}= L_{s~max} =  \frac { \ell_s}{2^{1/4}}= 0.841~ \ell_s  
\label{eq:rgamss}
\end{equation}
\\
Eq.~(\ref{eq:rgs}) shows the relation between the  Schwarzschild radius and the cosmological constant (Eq.~(\ref{eq:Lan})) when $T_{sem~bhAdS} = T_s$ ( $bhAdS$ string regime). Also a black hole in $AdS$ space time allows also an intermediate string regime, not present in the Schwarzschild black hole alone ($(H=0)$), since in the $bhAdS$ background there are two characteristic string scales as well: $ L_s$ and $\ell_s$. 
In an asymptotically $AdS$ space time and when $T_{sem~bhAdS}$ reaches $T_s$ , the black hole radius $r_g$ becomes either $r_{g +}$ or $r_{g -}$, depending on the $AdS$ string size $L_s$. Then, when the $AdS$ characteristic length $L_{c\ell}$ reaches its minimal value, $r_g$ becomes uniquely determined by $\ell_s$, as given by Eq.(\ref{eq:rgamss}). In addition, the $bhAdS$ string regime determines a maximal value for $H$ :  $H_{max} = 0.841~c/\ell_s$

Finally, the $bhAdS$ string regime is larger than black hole de Sitter (bhdS) string regime. Notice the differences between the bhAdS and bhdS string regimes~\cite{15}: 

(i) Only one root ($r_{g-}$) is present in the bhdS string regime. 

(ii) There is no condition such as Eq.(\ref{eq:rgc}) or (\ref{eq:rgams}) for $L_{c\ell}$ or $L_s$ in the bhdS string regime. 

ii) In the $bhdS$ space time, when $T_{sem~bhdS}$ reaches $T_s$, the black hole radius $r_g$ becomes $L_s$. Then, when $L_{c\ell} = \ell_s$, $r_g$ is minimal and determined by $\ell_s$ too,  ($r_{g~min} = 0.365~\ell_s$).
 
In contrast, in the $bhAdS$ background, $L_{c\ell}$ cannot reach $\ell_s$, $L_{c\ell~min}$ is  larger than $\ell_s$ (Eq.(\ref{eq:rgc})). The minimal black hole radius in $AdS$ space-time, $r_{g~ min} = 0.841~  \ell_s$, is {\bf larger} than the minimal black hole radius in de Sitter space: 
\begin{equation}
r_{g~min~bhAdS} = 2.304~ r_{g~min~bhdS}
\label{eq:rminAdS}
\end{equation}\bigskip

5.2.4. \emph{In Kerr- Newman black hole space time}\bigskip

For Kerr-Newman background the canonical partition function is: 
\begin{equation}
\ln Z = \frac{V_{D-1}}{(2 \pi)^{D-1}} \sum_{j, \alpha} \int^{\infty}_{m_0} d \Big( \frac{m}{m_s}\Big) \rho_s(m, j, q) 
\int d^{(D-1)}k 
 \ln \Bigg\{\frac{1 + \exp \Big\{-\beta_{sem} [(m^2 c^4 + \hbar^2 k^2 c^2)^{\frac{1}{2}}-\mu_{\alpha}]\Big\}}
{1 - \exp \Big\{-\beta_{sem} [(m^2 c^4 + \hbar^2 k^2 c^2)^{\frac{1}{2}}-\mu_{\alpha}]\Big\}}\Bigg\} 
\label{eq:Z}
\end{equation}
where $\rho_s(m, j, q)$ is given by Eqs.~(\ref{eq:roF}) and (\ref{eq:romsq}), here $T_{sem}\equiv T_{sem}(J, Q)$ (Eq.~(\ref{eq:TsemJQ})) and $\beta_{sem}=(k_B T_{sem}(J, Q))^{-1}$; $\mu_{\alpha}$,  $q_{\alpha}$ and $j_{\alpha}$ and are the chemical potential, charge, and angular momentum (about the axis of rotation of the black hole: $j = n_{\alpha}\hbar$) of a string mode $\alpha$ respectively; $m_0$  is the lowest string mass for which the asymptotic string density of mass level is valid, and $m_{s}$ is the fundamental string mass scale (Eq.~(\ref{eq:ms})).  

The chemical potential $\mu_{\alpha}$ is given by
\begin{equation}
\mu_{\alpha} = j_{\alpha}\Omega + q_{\alpha}\Phi 
\label{eq:muu}
\end{equation}
where $\Omega$ and $\Phi$ are the Kerr-Newman angular velocity and electric potential respectively~\cite{16}:
\begin{equation}
\Omega = \frac{c L_J}{r_+^2 + L^2_J}, ~~~~
\Phi = \frac{Q  r_+}{r_+^2 + L^2_J}
\label{eq:OFI}
\end{equation}

From Eq.~(\ref{eq:Z}) we have
\begin{equation}
\ln Z = \frac{4 V_{D-1}}{(2 \pi)^{D/2}} \frac{c}{\beta_{sem}^{\frac{D-2}{2}}\hbar^{(D-1)}}  
\sum_{j, \alpha} \sum_{n=1}^{\infty} \frac{e^{(2n-1) \beta_{sem} \mu_{\alpha}}}{(2n-1)^{D/2}}  
\int^{\infty}_{m_0} d \Big(\frac{m}{m_s}\Big) \rho_s (m, j, q)  m^{D/2} K_{D/2} \big[ \beta_{sem} (2n-1) m c^2 \big]
\label{eq:lnZ}
\end{equation}
being $K_{D/2}$ the modified Bessel function. Considering the leading order ($\beta_{sem}~ mc^2 \gg1$) and the asymptotic flat behavior of the string mass density of states, the leading contribution for the partition function is
\begin{equation}
\ln Z \simeq \frac{2 V_{D-1}} {(2 \pi)^{\frac{D-1}{2}}} \frac{m_s^{a-1}} { (\beta_{sem} \hbar^2)^{\frac{D-1}{2}}}  
\sum_{\alpha} e^{\beta_{sem} \mu_{\alpha}}
\int^{\infty}_{m_0} dm~ m^{(-a + \frac{D-1}{2})}~ e^{-( \beta_{sem}-\beta_{s}) mc^2}
\label{eq:Zaprox}
\end{equation}
where $\beta_{s} = (k_B t_{s})^{-1}$ (Eq.~(\ref{eq:ts})). As already mentioned, the factor $2$ in front of Eq.~(\ref{eq:Zaprox}) stands for both bosonic and fermionic strings included (otherwise, this factor is absent for either bosonic or fermionic strings).
Eq.~(\ref{eq:Zaprox}) implies that the black hole temperature $T_{sem}\equiv T_{sem}(J,Q)$ is bounded by the string temperature $t_{s}$: $T_{sem}(J, Q) \leq t_{s}$. 

 Finally, for $T_{sem}(J, Q) \ll t_{s}$ ($\beta_{sem} \gg \beta_{s}$) and $T_{sem}(J, Q) \rightarrow t_{s}$ ($\beta_{sem} \rightarrow \beta_{s}$), the partition function behaviors are respectively
\begin{equation}
\ln Z \simeq V_{D-1} \Big( \frac{m_s}{2 \pi \beta_{sem} \hbar^2} \Big)^{\frac{D-1}{2}} 
\sum_{\alpha} e^{\beta_{sem} \mu_{\alpha}}~~ e^{-\beta_{sem}m_0 c^2}
\label{eq:TTs}
\end{equation}
and
\begin{equation}
\ln Z \simeq V_{D-1} \Big( \frac{m_s}{2 \pi \beta_{s} \hbar^2} \Big)^{\frac{D-1}{2}} 
\sum_{\alpha} e^{\beta_{sem} \mu_{\alpha}} \frac{1} {\Big[(\beta_{sem} - \beta_{s})m_s c^2 \Big]}
\label{eq:Zo}
\end{equation}
analogously to Eqs.~(\ref{eq:ZTh}) and~(\ref{eq:ZmMs}), but here with the semiclassical temperature $T_{sem}(J, Q)$.  Eq.~(\ref{eq:Zo}) shows an universal pole singularity for any $D$ at the temperature $t_s$, typical of a string system with intrinsic Hagedorn temperature, and indicates a string phase transition of Carlitz's type~\cite{33} to a condensate finite energy state. i.e the transition takes place at $T_{sem}(J, Q)$ = $t_{s}$ towards a microscopic finite energy condensate of size range $l_{s}$. This stringy state forms, at the last stage of black hole evaporation, from the massive very excited strings emitted by the black hole, as we will show from the string emission cross section computed in Sec. 6 below.

We have just seen, from the canonical partition function Eq.~(\ref{eq:Zaprox}),  that the Kerr Newman black hole temperature (Eqs.~(\ref{eq:Tsem}) and~(\ref{eq:LJQ})) is bounded by the string temperature $t_{s}$ (Eq.~(\ref{eq:ts})). 
Therefore, the  string bound 
\begin{equation}
T_{sem}(J, Q) \leq t_{s}
\label{eq:Tsb}
\end{equation}
implies
\begin{equation}
L_{c\ell}(J, Q) \geq \hat{l}_{s}.
\label{eq:Lsb}
\end{equation}
Eqs.~(\ref{eq:LJQ}) and~(\ref{eq:ts}) yield
\begin{equation}
1~-~\frac{\nu^{2}}{2}\geq \delta~\sigma,~~~~~~~~
\sigma \equiv \frac{\hat{l}_{s}}{r_g}~-~1
\label{eq:nub}
\end{equation}
The above relations (Eqs.~(\ref{eq:Lsb}) and~(\ref{eq:nub}) lead to three different situations~\cite{12}:

(i) $ \sigma < 1$ (i.e. $r_g > \hat{l}_{s}/2$). In this case, the inequality  Eq.~(\ref{eq:nub}) is satisfied for all nonvanishing values of $\mu$ and $\nu$. 

Then $T_{sem}(J,Q) < t_{s}$ is always verified without restrictions on $J$ and $Q$.

(ii) $ \sigma = 1$ (i.e. $r_g = \hat{l}_{s}/2$). One has $\mu=0$ = $\nu$ (with $L_{c\ell}(J=0=Q)\equiv 2r_g=~ \hat{l}_{s}$, Eqs.~(\ref{eq:LJQ}) and~(\ref{eq:ts}) for the equal sign in Eq.~(\ref{eq:nub}); and $\mu\neq 0$, $\nu\neq 0$ or ($\mu\neq 0$, $\nu$= $0$), or ($\mu = 0$, $\nu\neq 0$) for the strict inequality. 

 Then $T_{sem}(J,Q) < t_{s}$ with any value of $J$ and $Q$ excluding both J and Q simultaneously equal to zero. The string limit $T_{sem}(J,Q) = t_{s}$ is reached with both $J = 0$ and $Q = 0$. 

(iii) $ \sigma > 1$ (i.e. $r_g < \hat{l}_{s}/2$). In this case, two critical values, $ \mu_{0}$  and $\nu_{0}$, appear for the angular momentum parameter $\mu$ and the charge parameter $\nu$ respectively. They are given by
\begin{equation}
\mu_{0}^{2} = \frac{4~\left(\sigma^{2}-1\right)\left(1-\nu^{2}\right)~-~\nu^{4}}{4~\sigma^{2}}
\label{eq:muo}
\end{equation}
and
\begin{equation}
\nu_{0}^{2} = 2~\left( 1~-~\sigma^{2}~+~\sigma\sqrt{\sigma^{2}~-~1}\right)
\label{eq:nuo}
\end{equation}

If $\nu\geq\nu_{0}$, Eq.~(\ref{eq:nub}) is fulfilled for all $\mu's$ and saturated for $\nu=\nu_{0}$ and $\mu=0$. In the opposite case i.e. $\nu<\nu_{0}$, one must have $\mu\geq\mu_{0}$. Here, $L_{c\ell}(J,Q) = \hat{l}_{s}$ if $\mu = \mu_{0}$.

In this case, there  exist a critical value for the charge, $Q_{0}$, and a critical value for the angular momentum, $J_{o}$:  For $Q\geq Q_{0}$, $T_{sem}(J, Q)\leq t_{s}$ holds for all $J$; the string limit $T_{sem}(J, Q)$ = $t_{s}$ is reached for $Q =Q_{0}$ and $J=0$, and $Q_{0}$ is given by
\begin{equation}
	Q_{0}^{2} =  \frac{G}{2}
	                            \left( 4 M^{2}~-~\left( M_{s} - 2M \right)^{2}
	                                    + \left( M_{s} - 2M\right) \sqrt{ \left(M_{s}~-~2M\right)^{2} - 4M^{2}} \right)
\label{eq:Q0}
\end{equation}

Otherwise, namely if $Q < Q_{o}$, there is a minimal angular momentum, $ J\geq J_{o}$, given by:
\begin{equation}
	J_{0}^{2} =  \frac{ \frac{ G^{2} M^{4} }{ c^{2} }
			             \left( \left(M_{s}-2M\right)^{2} - 4M^{2} \right)
			             \left(1-\frac{ Q^{2} }{ GM^{2} }\right)
			             - \frac{ Q^{4}~M^{2} }{ c^{2} }                                                                   
			         }{\left(M_{s}~-~2M\right)^{2}}
\label{eq:J0}
\end{equation}
($M_s\equiv (c^2/G)~\hat{l_s}$). Here, the limit $T_{sem}(J,Q) = t_{s}$ is reached for $J =J_{0}$.

From this analysis we can conclude that  -  given a semiclassical Kerr Newman black hole with mass $M$, angular momentum $J$, charge $Q$ and temperature $T_{sem}(J,Q)~<~ t_{s}$  - there are three possible cases for its evolution into a string state with temperature $T_{s}$:

(I) If $r_g > \frac{ \hat{l}_s}{2}$, $J\neq 0$ and $Q\neq0$, the semiclassical Kerr Newman black hole reaches a string state of temperature $t_{s}$, mass $M = M_{s}/4$, angular momentum  $J = 0$ and charge $Q = 0$.

(II) If $ r_g < \frac{ \hat{l}_s}{2}$, $J\neq0$ and $Q > Q_{0}$ ($Q_{0}$ being the critical value Eq.~(\ref {eq:Q0})), the string temperature $t_{s}$ is reached with $J=0$ and $Q=Q_{0}$.

(III) If $r_g < \frac{ \hat{l}_s}{2}$, $J >J_{0}$ and $Q < Q_{0}$, the string temperature $t_{s}$ is reached for $J=J_{0}$, being $J_{0}$ the minimal value Eq.~(\ref {eq:J0}). In this string state both angular momentum and charge are different from zero. 

The final conclusion is that, besides the classical upper bounds implying a maximal angular momentum and maximal charge for the black hole (Eq.~(\ref {eq:bcl})), there are {\it minimal} values for the Kerr Newman black hole angular momentum and the charge in the string regime.
This implies also minimal values for the angular momentum of a Kerr Black Hole, and for the charge of a Reissner-Nordstrom Black Hole. \bigskip

{\bf 6. QUANTUM STRING EMISSION BY A BLACK HOLE AND THE LAST STAGE OF BLACK HOLE EVAPORATION} \bigskip

As cosmological evolution goes from quantum string (QS) phase to quantum field theory (QFT) phase (and then to the classical epoch), black hole evaporation goes from a QFT phase to a QS phase. We will analyzed in this section the quantum string emission of a Schwarzschild black hole ($J=0$, $Q=0$) in an asymptotically flat, $dS$ and $AdS$ space times; and the quantum string emission of a rotating charged black hole ($KNBH$). Finally, we will consider the black hole quantum decay. 

For a non rotating uncharged black hole in a flat, $dS$ or $AdS$ space times, evaporation is measured by an observer which is at the corresponding asymptotic region i.e flat, $dS$ or $AdS$ backgrounds.

The quantum field emission cross section $\sigma_{QFT} (k) $ of a given emitted species of particles in a mode $k$ by a black hole in a given background is~\cite{37} -\cite{38},~\cite{59}
given by
\begin{equation}
\sigma_{QFT}(k)=\frac{\Gamma_A}{e^{(\beta_{sem} E(k))}-1}
\label{eq:sig}
\end{equation}
where $\Gamma_A$ is the greybody factor (absorption cross section), and we consider only the isotropic term (i.e. oscillatory behavior as a function of k is disregarded)~\cite{60}. For the sake of simplicity, only bosonic states have been considered ($+$ sign in the denominator for fermionic)  ; $\beta_{sem}=( k_B T_{sem} )^{-1}$, and $T_{sem}$ is the semiclassical - Hawking temperature in the chosen background. The quantum field emission cross section of particles of mass $m$ is defined as
\begin{equation}
\sigma_{QFT}(m) = \int_{0}^{\infty} \sigma_{QFT} (k)~ d\mu (k) 
\label{eq:sm}
\end{equation}
where $d\mu (k)$ is the number of states between $k$ and $k+dk$:
\begin{equation}
d\mu(k) = \frac{2 V_{D-1}}{\Big(4\pi\Big)^{\frac{D-1}{2}}\Gamma \Big( \frac{D-1}{2} \Big) }~k^{D-2}~dk
\label{eq:mu}
\end{equation}

From Eq.~(\ref{eq:sm}) we have
$$ \sigma_{QFT}(m)=\frac{V_{D-1}~\Gamma_A}{(2 \pi)^{\frac{D-1}{2}}}
\frac{\Big( mc^2\Big)^{\frac{D-2}{2}}~}{(\beta_{sem})^{D/2}~ (\hbar c)^{D-1}} ~~
\times $$
\begin{equation}
\sqrt{\frac{2}{\pi}}~\sum_{n=1}^{\infty} \frac{1}{n^{D/2}}~
\Big\{ n\beta_{sem} mc^2 K_{_{D/2}} (n\beta_{sem}mc^2) + K_{_{D/2 - 1}}  (n\beta_{sem} mc^2) \Big\} 
\label{eq:smD}
\end{equation}

For large $m$ and the leading order $n=1$ $(\beta_{sem}~ mc^2\gg1)$,  we have
\begin{equation}
\sigma_{QFT}(m) \simeq \frac{V_{D-1}~\Gamma_A} {(2\pi )^\frac{D-1}{2}}~ ~
\frac{m^{\frac{D-1}{2}}} {\left(\beta_{sem} ~\hbar^{2}\right)^ \frac{D-1}{2}}~ e^{-\beta_{sem}~ mc^2}
\label{eq:smDl}
\end{equation}

The string quantum emission cross section, $\sigma_{string}$, will be given by
\begin{equation}
\sigma_{string} \simeq  \int_{m_0}^{M_s} \rho_{s}(m)~\sigma_{QFT}(m)~ d\Big(\frac{m}{m_s}\Big)  
\label{eq:sDs}
\end{equation}
where $\rho_{s}(m)$ is the asymptotic density of mass levels in flat, $dS$ and $AdS$ space times (Eqs.~(\ref{eq:roa}), (\ref{eq:rob}), (\ref{eq:roc})).\bigskip

{\bf 6.1. Schwarzschild black hole quantum emission}\bigskip 

For a Schwarzschild black hole in an asymptotically flat space time, the black hole evaporation will be measured by an observer at this asymptotic region. Inserting the flat mass density of states  (Eq.~(\ref{eq:roa})) in Eq.~(\ref{eq:sDs}), we have the following behaviors for low and high black hole temperatures compared with the string temperature $t_s$ (string spin considerations are overlooked her; emission is larger for spinless particles~\cite{61}) :

For low temperatures, $T_{sem} \ll t_s$, we recover the semiclassical (QFT) Hawking emission at the temperature $T_{sem}$ (Eq.~(\ref{eq:LBSA})) 
\begin{equation}
\sigma_{string}\simeq \frac{V_{D-1}~\Gamma_A}{(2 \pi)^{\frac{D-1}{2}}}~
\frac{m_s^{\frac{D-3}{2}}}{\beta_{sem}~^{\frac{D+1}{2}}~ (\hbar c)^{D-1}}~e^{-\beta_{sem}m_0c^2}
\label{eq:sHaw}
\end{equation}

But for $T_{sem}\rightarrow t_s$, we have a singular pole behavior at $t_s$ (for any $D$ space time dimensions)
\begin{equation}
\sigma_{string} \sim \frac{V_{D-1}~\Gamma_A} {(2\pi )^\frac{D-1}{2}}~
\frac{ ~m_s^{\frac{D-3}{2}}} {\left(\beta_{sem}~\hbar^{2}\right)^{\frac{D-1}{2}}}~
\frac{ e^{-(\beta_{sem}-\beta_{s})~m_0 c^2}}{\Big(\beta_{sem} -\beta_{s}\Big)c^2}
 \label{eq:smia}
\end{equation}

We see again that a phase transition takes place~\cite{62}, at the string temperature $t_s$ (Eq.~(\ref{eq:ts})), towards a microscopic finite energy condensate of size $\hat{l_s}$~\cite{11}. This stringy state forms, at the last stage of black hole evaporation, from the massive very excited strings emitted by the black hole. The phase transition undergone by the emitted strings represents the non perturbative back reaction effect of the string emission on the black hole. As we already mentioned, an explicit dynamical perturbative solution to the back reaction effect of the string emission on the Schwarzschild black hole accompasses this picture: the black hole losses its mass, reduces its radius until $\hat{l}_s$ and rises its temperature until $t_s$.~\cite{11}.\bigskip

{\bf 6.2. Black hole - de Sitter quantum emission}\bigskip 

For $m \ll M_s$, (away from the upper mass bound and the temperature $T_s$), the $\rho_s(m, H)$ leading behavior is given by the flat space solution ($H=0$; Eq.~(\ref{eq:roa})), and the black hole emission coincides with Eq.~(\ref{eq:smia}) where $\beta_{sem}$ is substituted by $\beta_{sem~bhdS}$ and $t_s$ by $T_s$. This means that, for low $Hm$ regime, the string emission cross section shows the same singular behavior near $t_s$ as the low $Hm$ behavior of the canonical $dS$ partition function, and as the quantum string emission by a (asymptotically flat) black hole which was the previous case.
This is so, since in the $bhdS$ background, the string mass scale for low string masses (temperatures) is the Hagedorn (flat space) string temperature $t_s$; the limit $T_{sem~bhdS} \rightarrow t_s$ is a high temperature behavior for low $Hm \ll c/\alpha'$, and $t_s$ is smaller than the string $dS$ temperature $T_s$.

For low temperatures $\beta_{sem~bhdS} \gg \beta_{s}$ we recover the semiclassical (QFT) Hawking emission (Eq.~(\ref{eq:sHaw})) but here at the temperature  $T_{sem~bhdS}$. 
 
 However for high masses ($m \sim M_s$), i.e close to the string critical temperature $T_S$, the quantum emission cross section has a different behavior as the one of the previous case. We have for the $\sigma_{string}$ leading behavior :
\begin{equation}
\sigma_{string} ~~(T \sim T_s) \sim V_{D-1}~\Gamma_A~\left(\frac{k_B T_{sem~bhdS}}{\hbar c}\right)^{D-1}
\sqrt{1~-~\frac{T_{sem~bhdS}}{T_s}}
\label{eq:siT3}
\end{equation}

The black hole-de Sitter emission cross section shows a {\bf phase transition} at $T_{sem~bhdS} = T_{s}$: the string emission by the black hole condensates into a de Sitter string state of string de Sitter temperature $T_s$~\cite{14}. This is not like the flat (or asymptotically flat) space string phase transition (of Carlitz type), but this is a de Sitter type transition. Instead of featuring a single pole singularity in $(~T~ -~T_s~)$, the  transition is a square root branch point. The branch point singular behavior at $T_{s}$ is valid for any D-dimensions and is like the one we found for the de Sitter canonical partition function (Eq.~(\ref{eq:ZhT})) and for the de Sitter microscopic string density of states $\rho_s(m, H) $ Eq.~(\ref{eq:rob}) in the high $m$ (high $Hm \rightarrow c/\alpha'$) regime.  \bigskip

{\bf 6.3. Black hole - Anti de Sitter quantum emission}\bigskip 

We consider the quantum emission for the different mass ranges, with respect to the relevant $AdS$ string mass scale $M_s$  (Eq.~(\ref{eq:Ms})). For low masses $m \ll M_s$, the $\rho_s(m, |\Lambda|)$ leading behavior is given by the flat space solution. 
($H=0$; Eq.~(\ref{eq:roa})), and the black hole emission coincides with Eq.~(\ref{eq:smia}) where $\beta_{sem}$ is substituted by $\beta_{sem~bhAdS}$ and $t_s$ by $T_s$ (Eq.~(\ref{eq:Ts})).  
In other words, for a low regime, $|H|m \ll c/\alpha'$ regime, the $bhAdS$ string emission cross section shows the same singular behavior near $t_s$ as the low $|H|m$ behavior of the $dS$ and $AdS$ partition functions (Eq.~(\ref{eq:ZmMs})), and as the quantum string emission by an asymptotically flat black hole ~\cite{1},~\cite{4}, but here at the $bhAdS$ temperature $T_{sem~bhAdS}$. This is so, since in the $bhAdS$ background, the string mass scale for low string masses (temperatures) is the Hagedorn flat space string temperature $t_s$. The limit $T_{sem~bhAdS} \rightarrow t_s$ is a high temperature behavior for low $Hm \ll c/\alpha'$; $t_s$ is larger than $T_{sem~bhAdS}$ but smaller than the $AdS$ string temperature $T_s$.

For low temperatures $\beta_{sem~bhAdS} \gg \beta_{s}$, (i.e. semiclassical regime), we recover the QFT Hawking emission at the temperature  $T_{sem~bhAdS}$.

Now, for high masses  $m\sim M_s$, and $m >> M_s$, $\sigma_{string}$ behaves as:
\begin{equation}
\sigma_{string} (m \sim M_s~~\text{and}~~m\gg M_s) \sim \frac{V_{D-1}~\Gamma_A~}{\left(\beta_{sem~bhAdS}~\hbar c\right)^{D-1}}~
\left(\frac{\beta_s}{\beta_{s~AdS}}\right)^{\frac{3-D}{4}}~~
e^{\frac{1}{\beta_{sAdS}}( \beta_s ~-~2\pi \beta_{sem~bhAdS})}~
 \label{eq:sigma3}
\end{equation}
and no phase transition occurs as we already knew~\cite{15}.\bigskip

{\bf 6.4. Kerr Newman quantum emission}\bigskip 

The quantum string emission by a Kerr Newman black hole is given by the cross section
\begin{equation}
\sigma_{string} = \sum_{j} \sum_{\alpha} \int_{m_0}^{\infty} \rho_{s}(m, j, \alpha)~\sigma_{\alpha}(m,j)~  
d\big( \frac{m}{m_s}\big) 
\label{eq:sD}
\end{equation}
where 
\begin{equation}
\sigma_{\alpha}(m, j) = \int_{0}^{\infty} \sigma_{\alpha}(k,  j)~ d\mu(k, j)
\label{eq:sDj}
\end{equation}
$\sigma_{\alpha}(k,  j)$ is the QFT  
emission cross section of particles of species ${\alpha}$ in a mode of frequency $k$, spin $j_{\alpha}$ ($= n_{\alpha}\hbar$), and charge $q_{\alpha}$:
\begin{equation}
\sigma_{\alpha}(k, j) = \frac{\Gamma_{\alpha}}{\exp \Big\{ \beta_{sem}[E(k) - \mu_{\alpha}] \Big\} + (-1)^{2j +1}}
\label{eq:sk}
\end{equation}
where $\beta_{sem}= (k_{B}T_{sem}(J, Q))^{-1}$. $\Gamma_{\alpha}$ is the classical absorption cross section (grey body factor), and $\mu_{\alpha}$ the chemical potential given by Eq.~(\ref{eq:muu}).

The QFT  emission cross section of particles of mass $m$ and spin $j$ is defined as
\begin{equation}
\sigma(m) = \sum_{j} \sum_{\alpha} \int_{0}^{\infty} \sigma_{\alpha}(k, j)~  d\mu(k, j) 
\label{eq:smj}
\end{equation}
where $d\mu(k,j)$ is the number of states between $k$ and $k+dk$
\begin{equation}
d\mu(k, j) = n_{j} \frac{2 V_{D-1}}{(4 \pi)^{\frac{D-1}{2}}}~\frac{k^{D-2}}{\Gamma \Big( \frac{D-1}{2} \Big)} ~dk
\label{eq:muj}
\end{equation}
and $n_{j}$ is the number of spin states: $[( j+ D -3)! (2j + D -2)]~ [j! (D- 2)!]^{-1}$ for $SO(D)$; ( $2j+1$ for $SO(3)$).

From Eq.~(\ref{eq:smj}) we have
\begin{eqnarray}
&&\sigma(m)=\sum_{j} \sum_{\alpha}n_{j} \frac{V_{D-1}}{(2 \pi)^{\frac{D-1}{2}}}
\frac{\Gamma_{\alpha}~~\Big( mc^2\Big)^{\frac{D-2}{2}}}{\beta_{sem}^{D/2}~ (\hbar c)^{D-1}} ~~\sqrt{\frac{2}{\pi}}~\sum_{n=1}^{\infty} \frac{(-1)^{(n-1)2j}}{n^{D/2}}
\times \nonumber \\
&& e^{n \mu_{\alpha}\beta}
\Big\{n\beta_{sem}mc^2K_{_{D/2}}(n\beta_{sem} mc^2)+K_{_{D/2-1}}  (n\beta_{sem}mc^2) \Big\} 
\label{eq:smDr}
\end{eqnarray}
$K_{_{D/2}}$ being the modified Bessel function. For large $m$ and leading order $(n=1)$, Eq.~(\ref{eq:smDr}) becomes
\begin{equation}
\sigma(m)\simeq \sum_{j} \sum_{\alpha}n_{j}\frac{V_{D-1}}{(2 \pi)^{\frac{D-1}{2}}}
\frac{ \Gamma_{\alpha}~~\Big( mc^2\Big)^{\frac{D-3}{2}} }{\beta_{sem}^{\frac{D+1}{2}}~ (\hbar c)^{D-1}}
e^{-\beta_{sem}(mc^2-\mu_{\alpha})}
\label{eq:smDj}
\end{equation}
Therefore, from the quantum emission of strings by the black hole $\sigma_{string}$ (Eq.~(\ref{eq:sD})), we have the following leading behaviors:

For $\beta_{sem}\gg \beta_{s}$ 
\begin{equation}
\sigma_{string}\simeq \sum_{j} \sum_{\alpha}n_{j}\frac{V_{D-1}}{(2 \pi)^{\frac{D-1}{2}}}~
\frac{\Gamma_{\alpha}~~m_s^{\frac{D-3}{2}}}{\beta_{sem}^{\frac{D+1}{2}}~ (\hbar c)^{D-1}}~
e^{\beta_{sem}~ \mu_{\alpha}}~e^{-\beta_{sem}~m_0c^2}
\label{eq:sKNls}
\end{equation}

and for $\beta_{sem} \rightarrow \beta_{s}$ 
\begin{equation}
\sigma_{string}\simeq \sum_{j} \sum_{\alpha}n_{j}\frac{V_{D-1}}{(2 \pi)^{\frac{D-1}{2}}}~
\frac{\Gamma_{\alpha}~~m_s^{\frac{D-3}{2}}}{\beta_{s}^{\frac{D-1}{2}}~ (\hbar c)^{D-1}}~
e^{\beta_{s} \mu_{\alpha}} \frac{1}{(\beta_{sem} - \beta_s)}
\label{eq:sDTs}
\end{equation}

The string emission cross section shows that for $T_{sem}(J, Q)\ll t_{s}$ the emission is thermal with temperature $T_{sem}(J, Q)$ (Eq.~(\ref{eq:LJQ})), this is the Hawking part of the emission in the early stage of evaporation, that is, the semiclassical or QFT regime. As evaporation proceeds, $T_{sem}(J, Q)$ increases:
for $T_{sem}(J, Q)\rightarrow t_{s}$, the massive string modes dominate the emission; the string emission cross section shows at $t_{s}$ the same behavior as ln$Z$ (Eq.~(\ref{eq:Zo})), that is, a phase transition takes place at $T_{sem}(J, Q)$ = $t_{s}$. This phase transition undergone by the emitted strings represents the non-perturbative back reaction effect of the string emission on the black hole.
In a non-singular finite process, the temperature does not becomes infinite but remains bounded by $t_{s}$ (the radius and mass do not reduce to zero)~\cite{12}.\bigskip

{\bf 6.5. Quantum Black hole decay}\bigskip 

In the semiclassical (QFT) regime, ie in the early stages, of black hole evaporation, the black holes $BH$, $bhdS$, $bhAdS$, $KBH$, $RNBH$ and $KN$ decay as a grey body at the Hawking temperature $T_{sem}$, with a decay rate
\begin{equation}
\Gamma_{sem} = \left| \frac{ d\ln M_{sem} }{ d t}\right| \sim G~T_{sem}^{3}~~,~~M_{sem}= 8~\pi~ T_{sem}~ (\hbar=c=k_{B}=1).           
\label{eq:decay}
\end{equation}
where $T_{sem}$ (Eq.(\ref{eq:Tsem})) is given by Eqs.(\ref{eq:LBSA}), (\ref{eq:LbhSA}), (\ref{eq:LBH}) and~(\ref{eq:LJQ}) respectively.
As evaporation proceeds, $T_{sem}$ increases until it reaches the string temperature $T_{string}$, where 
$T_{string}=t_s$ (Eq.(\ref{eq:ts})) for $BH$, $KBH$, $RNBH$ and $KN$ black holes and $T_{string}=T_s$ (Eq.(\ref{eq:Ts})) for $bhdS$ and $bhAdS$ black holes. 
The black hole itself becomes a very excited string state in any of the backgrounds considered. This quantum string state decays in the usual way quantum strings do (~\cite{63},\cite{64}), ie with a width, 
\begin{equation}
\Gamma_{s}\sim \alpha'~T_{string}^{3}           
\label{eq:des}
\end{equation}
As $T_{sem} \rightarrow  T_{string}$, $\Gamma_{sem}$ becomes $\Gamma_{s}$ ($G\sim \alpha'$),  and the final decay is a pure (non mixed) quantum mechanical string decay into all type of particles. 
The string \emph{minimal} black hole will have a life time $\tau = ( \Gamma_{s})^{-1}$. In this \emph{effective} string framework there is no loss of information, i.e no paradox at all~\cite{29},~\cite{12},~\cite{14}-~\cite{15}.  \bigskip


{\bf 7. SEMICLASSICAL (Q.F.T) and QUANTUM (STRING) REGIMES}\bigskip

{\bf 7.1. Semiclassical entropy for (asymptotically flat) Schwarzschild black hole, de Sitter and Anti de Sitter space times}\bigskip

From our analysis in the previous Sections for different backgrounds, we have shown that for $T_{sem}\rightarrow T_{string}$, the semiclassical (Q.F.T) regime with Hawking-Gibbons temperature $T_{sem}$ undergoes a phase transition into a string phase at the string temperature $T_{string}$, being $T_{string}$ the critical temperature ($T_{string}=t_s$ (Eq.(\ref{eq:ts})) for $BH$, $KBH$, $RNBH$ and $KN$, and $T_{string}=T_s$ (Eq.(\ref{eq:Ts})) for $dS$,  $bhdS$ and $bhAdS$). This means that, in the quantum string regime, the semiclassical mass density of states $\rho_{sem}$ becomes the string mass density of states $\rho_s$, and the semiclassical entropy $S_{sem}$ becomes the string entropy $S_{s}$. Namely, in a given background, a semiclassical  state, $(dS)_{sem}  \equiv (T_{sem}, \rho_{sem}, S_{sem})$, undergoes a phase transition  into a quantum string state $(dS)_{s} \equiv (T_{string}, \rho_{s}, S_{s})$. 

The sets $(dS)_{s}$ and $(dS)_{sem}$ are the same quantities but in different (quantum and semiclassical/classical) regimes. This is the usual classical/quantum duality but in the gravity domain, which is {\it universal}, and not linked to any symmetry or isommetry nor to the number or the kind of dimensions. 
From the semiclassical $(dS)_{sem}$ and quantum string $(dS)_{s}$ regimes, we can write the full de Sitter entropy $S_{sem}$, with quantum corrections included, such that it becomes the string entropy $S_s(m)$ (Eq.~(\ref{eq:rhos})) in the string regime.

For the (asymptotically flat) $BH$, $dS$ and $AdS$ cases, the full semiclassical entropies $S_{sem}$ are given by~\cite{29},~\cite{12},~\cite{14}-~\cite{15}
 \begin{eqnarray}
S_{sem, flat}&\equiv& S_{sem}(M)= S_{sem}^{(0)} 
-a~k_B~\ln ~\big(\frac{S_{sem}^{(0)}}{k_B}\big)  
\label{eq:Sema}\\
S_{sem, dS}(H)&=& \hat {S_{sem}}^{(0)}(H) 
-a~k_B~\ln ~\big(\frac{\hat {S_{sem}}^{(0)}(H)}{k_B}\big) - k_B ~\ln \mathcal {F}(H)
\label{eq:Semb}\\
S_{s, AdS}(|\Lambda|)&=& \hat {S_s}^{(0)}(|\Lambda|) 
-a~k_B~\ln ~\big(\frac{\hat {S_{sem}}^{(0)}(|\Lambda|)}{k_B}\big) - k_B ~\ln \mathcal{F}(|\Lambda|)
\label{eq:Semc}
\end{eqnarray}
where for Eqs.~(\ref{eq:Semb}) and (\ref{eq:Semc}) 
\begin{equation}
\hat {S_{sem}}^{(0)}(m)\equiv S_{sem}^{(0)}\sqrt{g(X)}~~,~~\mathcal{F} \equiv \sqrt{(1 \mp 4X^2)g(X)}
\label{eq:SemsdsHF}
\end{equation}
($\mp$ for $dS$ and $AdS$ respectively) being $X$ the dimensionless variable 
\begin{equation}
2 X(|H|)\equiv \frac{\pi k_B}{S_{sem}^{(0)}(|H|)}=  \Big(\frac{m_{Pl}}{M_{cl}}\Big)^2   
\label{eq:Xem}
\end{equation}
($M_{cl}$ is the $dS(AdS)$ mass scale Eq.(\ref{eq:Mcl}), and $m_{Pl}$ the Planck mass Eq.(\ref{eq:mlP})) and
\begin{equation}
 g(X)\equiv \frac{2}{1+\Delta_{sem}},~~~~ \Delta_{sem} \equiv\sqrt{1\mp4X^2}= \sqrt{1 \mp\Big(\frac{\pi k_B}{S_{sem}^{(0)}(|H|)}\Big)^2} 
\label{eq:deltasem}
\end{equation}
($\mp$ for $dS$ and $AdS$ respectively). $S_{sem}^{(0)}$ and $S_{sem}^{(0)}(|H|)$ are the usual Bekenstein-Hawking entropies of flat, $dS$ ($H > 0$) and $AdS$ ($\Lambda < 0$) backgrounds (Eqs.~(\ref{eq:S0sem}) and~(\ref{eq:MBSA})).

From Eq.(\ref{eq:Sema}) we see that, for a Schwarzschild BH, the Bekenstein-Hawking entropy $S_{sem}^{(0)}$ is the leading term. Analogously, for low curvature regime in $dS$ ($AdS$) ( low $|H| \ll c/\ell_{Pl}$,  $M_{cl}\gg m_{Pl}$ i.e  $X \rightarrow 0$, $\Delta \rightarrow 1$,  $g(X)\rightarrow 1$) the Bekenstein-Hawking entropy $S_{sem}(|H|)$ is the leading term with its logarithmic correction: 
\begin{equation}
S_{sem}(|H|) = S_{sem}^{(0)}(|H|) -ak_B~\ln \Big(\frac{S_{sem}^{(0)}(|H|)}{k_B}\Big) 
\label{eq:SsemoH}
\end{equation}\bigskip

{\bf 7.2. Non extremal Kerr black hole entropy}\bigskip 

We can write the semiclassical entropy $S_{sem}(M, J)$ for the Kerr black hole such that it becomes the string entropy $S_s(m,j)$ in the string regime (Eq.~(\ref{eq:SjFs})), namely~\cite{12}:
\begin{equation}
S_{sem}(M, J) = S_{sem}^{(0)}(M, J) - a ~k_B~\ln \Big (~\frac{S_{sem}^{(0)}(M, J)}{k_B}~ \Big)~+~k_B~\ln~ F(S_{sem}^{(0)}, J)           
\label{eq:SsemBH}
\end{equation}
where $S_{sem}^{0}(M, J)$ is the Bekenstein-Hawking entropy (Eq.~(\ref{eq:S0sem}) and Eq.~(\ref{eq:MbhSA})) 
and $F(S_{sem}^{(0)}, J)$ is given by
\begin{equation}
F = \Delta^{-1} \Big(\frac{1 + \Delta}{2\Delta}\Big)^{a}~e^{\Big(\frac{1 -\Delta}{1 +\Delta}\Big) \frac{S_{sem}^{(0)}(M, J)}{\Delta k_B}} 
\frac{1}{\cosh^2 \Bigg(\frac{(1 - \Delta)}{\Delta}\frac{S_{sem}^{(0)}(M, J)}{2 k_B}\Bigg)} 
\label{eq:FBH}
\end{equation}

For $J=0$: $F =  1$ and $S_{sem}^{(0)}(M,J=0)\equiv S_{sem}^{(0)} = 4 \pi k_{B} \Big(M/m_{Pl}\Big)^2$, 
being $S_{sem}^{(0)}$ the Schwarzschild black hole Bekenstein-Hawking entropy. $\Delta$ is given by Eq.~(\ref{eq:landa}), which in terms of $S_{sem}^{(0)}$, reads:
\begin{equation}
\Delta = \sqrt{1 - \Big(\frac{J}{\hbar}\Big)^2 \Big(\frac{4 \pi k_B~}{S_{sem}^{(0)}} \Big)^2}
\label{eq:deltaBH}
\end{equation}
Therefore, the whole Kerr entropy $S_{sem}(M, J)$ (Eq.(~\ref{eq:SsemBH})) can be written in terms of $S_{sem}^{(0)}$:
\begin{equation}
S_{sem}(M, J) = \Big(\frac{1 + \Delta^2}{2 \Delta} \Big) S_{sem}^{(0)}~ -~ a~k_B~\ln \Big (~\frac{S_{sem}^{(0)}}{k_B}~ \Big) - 
(a + 1)~k_B~ \ln\Delta ~-~ 2~k_B~ \ln~\cosh \Big [~  \frac{S_{sem}^{(0)}}{4 k_B}  \frac{(1 - \Delta^2)}{\Delta}~ \Big] 
\label{eq:SJFcos}
\end{equation}

The first term in Eq.~(\ref{eq:SJFcos}) reads 
\begin{equation}
\Big[~1 - \frac {1}{2}~\Big(\frac{J}{\hbar}\Big)^2 \Big(~\frac{4 \pi k_B}{S_{sem}^{(0)}}~\Big)^2~
\Big]~ {\Delta}^{-1}~ S_{sem}^{(0)}.
\label{eq:factorS}
\end{equation}
For $\Delta \neq0$, the effect of the angular momentum is to reduce the entropy. $S_{sem}(M, J))$
is maximal for $J=0$ (ie for $\Delta = 1$). For $ J=0$, we reproduce the semiclassical entropy of a Schwarzschild black hole (Eq.~(\ref{eq:Sema})):
\begin{equation}
S_{sem}(M)~ = ~S_{sem}^{(0)}~-~ a ~k_B~\ln S_{sem}^{(0)}
\label{eq:Ssemzero}
\end{equation}
Notice that the  {\bf new} term  $\ln F( S_{sem}^{(0)}(M,J))$  in Eq.~(\ref{eq:SsemBH}) is enterely due to $J \neq 0$, and yields in particular to the last  $\ln cosh$  term in Eq.~(\ref{eq:SJFcos}). The argument of the last term in Eq.~(\ref{eq:SJFcos}) reads

\begin{equation}
X~\equiv~ \frac{S_{sem}^{(0)}}{4 k_B}\frac{(1 - \Delta^2)}{\Delta}~=~\frac{\pi}{\Delta}
\Big(\frac {J}{\hbar}\Big)^2\frac{4 \pi k_B}{S_s^{(0)}}~ =~ \frac{\pi}{\Delta}~ \Big(\frac{J}{\hbar}\Big)^2~ \Big(\frac{m_{Pl}}{M}\Big)^2
\label{eq:argX}
\end{equation}

Eq.~(\ref{eq:SsemBH}) provides the whole Kerr black hole entropy $S_{sem}(M,J)$ as a function of the Bekenstein-Hawking entropy $S_{sem}^{(0)}(M,J)$. 
For $M\gg m_{Pl}$  and  $J < GM^2/c$,  $S_{sem}^{(0)}$  is the leading term of this expression, {\bf but} for high angular momentum, (nearly extremal or extremal case $J= GM^2/c$), a gravitational {\bf phase transition} operates and the whole entropy $S_{sem}$ is drastically  {\bf different} from the Bekenstein-Hawking entropy $S_{sem}^{(0)}$, as we see below.\bigskip

{\bf 7.3. Extremal Kerr black hole and gravitational phase transition}\bigskip

The semiclassical extremal Kerr Newman black hole does not evaporate through Hawking radiation, as the Hawking temperature is zero in this case (Eqs.~(\ref{eq:TsemJQ}),~(\ref{eq:LJQ}) and~(\ref{eq:nuex})) 
\begin{equation}
T_{sem}(J, Q)_{extremal} = 0
\label{eq:Textremal } 
\end{equation}
Eq.~(\ref{eq:Tsb}) is a strict inequality in this case: $T_{sem}(J, Q)_{extremal} < t_s $. The string temperature cannot be reached (unless the extremal configuration would already be a stringy state).

The extremal black hole is, among the black hole states, the most stable configuration, i.e the most classical or semiclassical one. 

A Kerr-Newman black hole cannot become through quantum decay an extremal black hole. The extremal black hole cannot be the late state of black hole evaporation. Through evaporation and decay, the black hole losses charge and angular momentum (super-radiance like processes) at a higher rate than the loss of mass through thermal radiation. Thus, if a black hole was not extremal at its origin, it will not be extremal at its end. A Kerr Newman black hole evolves in general, through evaporation, into a Schwarzshild black hole, and becomes at its late stage a stringy state, which then decays, by the usual string decay process, in all types of massless and massive particles.

In particular, we consider now an extremal Kerr black hole. 
The Bekenstein-Hawking entropy $S_{sem}^{(0)}(M, J)$ is minimal in the extremal case, i.e when $J$ reaches its {\bf maximal} value ($\Delta = 0$, Eq.~(\ref{eq:landa}))~\cite{12}
\begin{equation}
S_{sem}^{(0)} (M, J)_{extremal} = \frac{1}{2}~S_{sem}^{(0)}(M, J=0) 
\label{eq:SJextremal}
\end{equation}

For $\Delta \rightarrow 0$, the last term in Eq.~(\ref{eq:SJFcos}) substracts the first one, and the pole in $\Delta$ of these two terms cancel out. $S_{sem}(M,J)_{extremal}$ is :
\begin{equation} 
S_{sem}(M,J)_{extremal} =   -(a + 1)~k_B~ \ln \frac {\Delta}{2}~ +~ k_B~\ln 2~ +~ \Delta ~\Big(~\frac{3} {4}~S_{sem}^{(0)}~-~ak_B~\Big)~+ ~ O(\Delta^{2})
\label{eq:Ssemextremal}
\end{equation}
In terms of the mass, or temperature, $\Delta$ (Eq.~(\ref{eq:landa})) is given by
\begin{equation} 
\Delta~ =~ \sqrt{1 - \Big(\frac{J}{\hbar}\Big)^2 \Big(\frac{m_{Pl}}{M} \Big)^4}~=~ \sqrt{1 - \Big(\frac{J}{\hbar}\Big)^2 \Big(\frac{T_{sem}}{T} \Big)^2}  
\label{eq:DeltaM}
\end{equation}
($ T = M c^2/8\pi k_B$) being now $T_{sem}$ the Schwarzschild Hawking temperature.
In the extreme limit $(J/\hbar) \rightarrow (M/m_{Pl})^2 $, $S_{sem}(M,J)_{extremal}$ is dominated by
\begin{equation}
S_{sem}(M,J)_{extremal} =  -(a + 1)~k_B~ \ln ~(\sqrt{\frac {2}{T}}\sqrt{T - \Big(\frac{J}{\hbar}\Big)^{1/2} {T_{Pl}}}~)~+~O(1)
\label{eq:SsemTextr}
\end{equation}
(being $T_{Pl}$ the Planck temperature). This shows that a {\bf phase transition} takes place at $T \rightarrow \sqrt{(J/\hbar)}~T_{Pl} $, we call it {\bf extremal transition}. 

The characteristic features of this gravitational transition can be discussed on the lines of the extremal string transition we analysed for the extremal string states (Subsec.4.1).\bigskip

{\bf 7.4. De Sitter gravitational phase transition}\bigskip 

For {\bf high} Hubble constant, $H \sim c/\ell_{Pl}$, (i.e $M_{cl}\sim m_{Pl}$), the high curvature or quantum $dS$ regime is very different from the $BH$, $dS$ low curvature and the high curvature $AdS$ regimes as we shall see. In the limit  $M_{cl}\sim m_{Pl}$, $S_{sem}^{(0)}(|H|)$ is subdominant for $dS$ background (and $AdS$ as well). In fact, 
for $\Delta_{dS} \sim 0$ ($\Delta_{sem}(dS)\equiv \Delta_{dS}$, Eq.~(\ref{eq:deltasem})), or $ M_{c\ell}\rightarrow m_{Pl}$, the entropy $S_{sem,dS}(H)$ (Eq.~(\ref{eq:Semb})) behaves as~\cite{14}:
\begin{equation}
S_{sem,dS}(H)_{\Delta_{dS} \sim 0} = k_B~ \ln \Delta_{dS} ~+~O(1)
\label{eq:SsemMcl}
\end{equation}
and the Bekenstein-Hawking entropy $S_{sem}^{(0)}(H)$ is sub-leading, O(1), ($S_{sem}^{(0)}(H)_{\Delta_{dS} = 0}  = \pi k_B$). 

In the limit $ M_{cl} \rightarrow m_{Pl} $,  $S_{sem,dS}(H)$ is dominated by
\begin{equation}
S_{sem,dS}(H)_{\Delta_{dS} \rightarrow 0} =  -~k_B~ \ln ~\Big(~\sqrt{2} \sqrt{1 - \frac{T_{semdS}}{T}}~\Big)~~+~O(1)
\label{eq:SsemPl}
\end{equation}
being
\begin{equation} 
\Delta_{dS}~ =~ \sqrt{1 -  \Big(\frac{m_{Pl}}{M_{cl}} \Big)^4}~=~ \sqrt{1 - \Big(\frac{T_{semdS}}{T} \Big)^2} 
\label{eq:DeltasemdS}
\end{equation}
($T = (1/2\pi k_B)~M_{cl} c^2$). This shows that a {\bf phase transition} takes place at $T \rightarrow T_{semdS}$, implying that the transition occurs for $M_{cl} \rightarrow m_{Pl}$, ie $T \rightarrow T_{Pl}$, (that is for high $H \rightarrow c/\ell_{Pl}$).  
This is a {\bf gravitational} like transition, similar to the de Sitter string transition we analysed in Sec. 4 : the signature of this transition is the square root {\it branch point} behavior at the critical mass (temperature) analogous to the thermal self-gravitating gas phase transition of point particles \cite {34}-~\cite{36},~\cite{65} and to the string gas in $dS$ space. This behavior is {\it universal}, and happens in any number of space-time dimensions~\cite{29},~\cite{66} . We have  already seen the same feature for the extremal Kerr black hole (high angular momentum $J\rightarrow M^2G/c$, extremal transition).\bigskip

{\bf 7.5. Absence of Anti de Sitter gravitational phase transition} \bigskip

For $AdS$ background, the variable $\Delta_{AdS}$ ($\Delta_{sem}(AdS)\equiv \Delta_{AdS}$) reads
\begin{equation} 
\Delta_{AdS}~ =~ \sqrt{1 +  \Big(\frac{m_{Pl}}{M_{cl}} \Big)^4}~=~ \sqrt{1 + \Big(\frac{T_{semAdS}}{T} \Big)^2} 
\label{eq:DeltasemAdS}
\end{equation}
($T = (1/2\pi k_B)~M_{cl} c^2$). It must be noticed that $\Delta_{AdS}$ {\it never} vanishes in this space-time, contrary to $dS$ space-time in which $\Delta_{dS} \rightarrow 0 $ (Eq.~(\ref{eq:DeltasemdS})) for high cosmological constant (high curvature or quantum $dS$ regime). For high curvature,  i.e. $M_{c\ell}\rightarrow m_{Pl}$, the full  $AdS$ entropy $S_{sem,AdS}(|\Lambda|)$ (Eq.~(\ref{eq:Semc})) is :
\begin{equation}
S_{sem,AdS}(|\Lambda| ) = \sqrt{\frac{2}{1+\sqrt{2}}}~ \pi k_B - ak_B~\ln \Big(\sqrt{\frac{2}{1+\sqrt{2}}}~\pi \Big) - k_B~\ln \Big(\frac{2}{\sqrt{1+\sqrt{2}}} \Big) 
\label{eq:SsemMcl}
\end{equation}
(Bekenstein-Hawking $AdS$ entropy: $S_{sem,AdS}^{(0)}(|\Lambda|)(M_{c\ell}= m_{Pl})= \pi k_B$ ). Contrary to $dS$,  {\bf no} phase transition occurs at $T \rightarrow T_{semAdS}$ in de $AdS$ space time.  That is, there is {\bf no} phase transition at $M_{cl} \rightarrow m_{Pl}$, ie $T \rightarrow t_{Pl}$, (high curvature $|\Lambda|^{1/2} \rightarrow c/\ell_{Pl}$ or quantum $AdS$ regime).  This is so, since for $AdS$ background, like for the $AdS$ string entropy, no singularity at finite mass or finite temperature occurs in the density of states or in the entropy.
Also, contrary to $dS$ and to extremal $KBH$ cases, there is not a square root branch point in the mass (temperature) analogous to the thermal self-gravitating gas phase transition of point particles.

It must be noticed that in $AdS$ background we can still analyze the mass regime  $M_{cl} << m_{Pl}$, for  which $ X >> 1$, $ \Delta_{AdS} = 2X >> 1$ and $ g(X) = 1/X  \rightarrow 0 $ (Eq.~(\ref{eq:Xem}) and~(\ref{eq:deltasem})). In this regime, the Bekenstein-Hawking term is subdominant:
\begin{equation}
S_{sem}^{(0)}(|\Lambda|)(X >> 1)= \frac {\pi k_B}{2X} << 1, 
\label{eq:Ssempl}
\end{equation}

therefore, $\sqrt {g(X)}S_{sem}^{(0)} = \pi k_B /( 2 X^{3/2})$ , and the full $AdS$ entropy $S_{sem,AdS}(|\Lambda|)$ (Eq.~(\ref{eq:Semc})) behaves as :
\begin{equation}
S_{sem,AdS}(|\Lambda|) (X >>1) = ak_B~\ln \Big(\frac{2}{\pi}~ X ^{3/2}\Big) - k_B~ \ln \Big (2X^{1/2}\Big) ~+~ \frac{\pi k_B}{2X^{3/2}}
\label{eq:SsmMmpl}
\end{equation}

This is a {\bf high} curvature quantum regime, in which $|\Lambda|^{1/2} >> c/\ell_{Pl}$, and therefore, the entropy is not dominated by the usual Bekenstein-Hawking (zero order term), which results negligable in this case~\cite{15}. The Planck scale  $AdS$ curvature regime ($|\Lambda|^{1/2} \sim c/\ell_{Pl}$), is reached {\bf smoothly}, without any singular behavior or phase transition in the entropy. The very high curvature $AdS $ regime, $|\Lambda|^{1/2} >> c/\ell_{Pl}$, is reached with a logarithmic growing behavior of the entropy. \bigskip

{\bf 8. SUMMARY AND CONCLUSIONS}\bigskip

In the framework of the \emph{effective} approach to quantum strings in curved backgrounds of physical relevance ($dS$, $AdS$, $BH$, $bhdS$, $bhAdS$, $KBH$, $RNBH$ and $KNBH$), explicit calculations of the quantum string entropy, string partition function and quantum emission by  black holes (Schwarzschild, rotating, charged, i.e. in an asymptotically flat, $dS$ and $AdS$ space times) have led to several {\bf new} results:

(i) For quantum strings, we have not only Hagedorn-Carlitz type of transitions, but {\bf new} gravitational phase transitions appear with a common distinctive {\bf universal} feature: a square root {\bf branch point} singularity in any space time dimensions. 

This gravitational phase transition is shown explicitly for $dS$ and $KNBH$ cases (also for quantum strings in flat space time when mode angular momenta are considered).  

The same behavior was already found for a thermal self-gravitating gas of point non relativistic particles (de Vega - S\'anchez transition), thus describing a new {\bf universality class}. 

(ii) On the contrary, there are not phase transitions for $AdS$ alone.

(iii) For $dS$ background, upper bounds for the Hubble constant $H$, and of the cosmological constant $\Lambda$, are found, dictated by the quantum string phase transition.

(iv) The last stage of black hole evaporation is a microscopic string state with a finite string critical temperature which decays as usual quantum strings do in pure quantum non-thermal radiation.  

Quantum black hole emission and black hole decay are described both in the semiclassical (Q.F.T) and quantum (string) regimes.

(v). In the $KNBH$ evaporation,  the black hole losses its mass and also its angular momentum and charge.

New lower string bounds are given for the Kerr-Newman black hole angular momentum and charge, which are entirely different from the upper classical and semiclassical bounds.

(vi). For Schwarzschild black holes in $dS$ space time, a relation between the Schwarzschild radius and the Hubble constant emerges, due to $bhdS$ phase transition.

(vii). For Schwarzschild black holes in $AdS$, a minimal classical length or a maximal string length are given. 

Furthermore, the minimal black hole radius in $AdS$ space time is larger than the minimal black hole radius in $dS$.

(viii). We have seen that, for a given background, a semiclassical  state, $(dS)_{sem}= (L_{classical}, T_{sem}, \rho_{sem}, S_{sem})$, undergoes a phase transition  into a quantum string state, $(dS)_{s}$ = $(L_{string}, T_{stringl}, \rho_{s}, S_{s})$. These sets, $(dS)_{s}$ and $(dS)_{sem}$, are the same quantities but in different (quantum and semiclassical/classical) regimes. This is considered as the usual classical/quantum (wave/particle) duality but in the gravity domain, which is {\it universal}, and not linked to any symmetry or isommetry nor to the number or the kind of space-time dimensions. 

(ix). From the full quantum string entropy $S_{s}$, we have written the semiclassical $S_{sem}$ entropy with quantum (logarithmic) corrections included. 

Gravitational phase transitions are shown for $dS$ and extremal $KBH$ cases. These transitions display the universal feature of a square root branch point singularity in the temperature.

(xi). For extremal $KBH$, a gravitational phase transition takes place at temperature greater than the Planck temperature. We call this \emph{extremal transition}. 

\begin{acknowledgements}
M. R. M. acknowledges the Spanish Ministry of Education and Science ( FPA 2004-2602 project) for financial
support, and the Observatoire de Paris, LERMA, for the kind hospitality extended to her.
\end{acknowledgements}


\end{document}